\documentclass[sigconf]{acmart}
\usepackage{makecell}
\usepackage{hyperref}
\usepackage{array}
\usepackage{booktabs}
\usepackage{tikz}
\usepackage{caption}
\usepackage{xspace}
\usepackage[normalem]{ulem}
\usepackage{tabularx}
\usepackage{csquotes}
\usepackage{ragged2e}

\newcolumntype{Y}{>{\raggedright\arraybackslash}X}

\AtBeginDocument{%
  }

\copyrightyear{2026}
\acmYear{2026}
\setcopyright{cc}
\setcctype{by}
\acmConference[CHI '26]{Proceedings of the 2026 CHI Conference on Human Factors in Computing Systems}{April 13--17, 2026}{Barcelona, Spain}
\acmBooktitle{Proceedings of the 2026 CHI Conference on Human Factors in Computing Systems (CHI '26), April 13--17, 2026, Barcelona, Spain}
\acmPrice{}
\acmDOI{10.1145/3772318.3791371}
\acmISBN{979-8-4007-2278-3/2026/04}

\newcommand{\systemName}{\textsc{WireWay}\xspace}

\newcommand\jw[1]{#1}

\renewcommand{\sout}[1]{}

\begin{document}

\title{Wire Your Way: Hardware-Contextualized Guidance and In-situ Tests for Personalized Circuit Prototyping}


\author{Punn Lertjaturaphat}
\authornote{Both authors contributed equally to this research.}
\affiliation{%
  \institution{KAIST}
  \department{Department of Industrial Design}
  \city{Daejeon}
  \country{Republic of Korea}
}
\email{punnlert@kaist.ac.kr}

\author{Jungwoo Rhee}
\authornotemark[1]
\orcid{0009-0005-3832-2835}
\affiliation{%
  \institution{KAIST}
  \department{Department of Industrial Design}
  \city{Daejeon}
  \country{Republic of Korea}
}
\email{jwoorhee@kaist.ac.kr}

\author{Jaewon You}
\affiliation{%
  \institution{KAIST}
  \department{Department of Industrial Design}
  \city{Daejeon}
  \country{Republic of Korea}
}
\email{ampersand328@kaist.ac.kr}

\author{Andrea Bianchi}
\orcid{0000-0002-7500-7974}
\affiliation{%
  \institution{KAIST}
  \department{Department of Industrial Design}
  \city{Daejeon}
  \country{Republic of Korea}
}
\affiliation{%
  \institution{KAIST}
  \department{School of Computing}
  \city{Daejeon}
  \country{Republic of Korea}
}
\email{andrea@kaist.ac.kr}

\renewcommand{\shortauthors}{Lertjaturaphat and Rhee et al.}

\begin{abstract}
The increasing popularity of microcontroller platforms like Arduino enables diverse end-user developers to participate in circuit prototyping. Traditionally, follow-along tutorials serve as an essential learning method for makers, and in fact, several prior toolkits leveraged this format as a way to engage new makers. However, literature and our formative study (N=12) show that makers have unique preferences regarding the construction of their circuits and idiosyncratic ways to assess and debug problems, which contrasts with the step-by-step instructional nature of tutorials and those systems leveraging this method. To address this mismatch, we present a prototyping platform that supports personalized circuit construction and debugging. Our system utilizes an augmented breadboard, which is circuit-aware and supports on-the-fly hardware reconfiguration via contextualized guidance and in-situ circuit validation through interactive tests. Through a usability study (N=12), we demonstrate how makers leverage circuit-aware guidance and debugging to support individual building patterns.
\end{abstract}


\begin{teaserfigure}
    \centering
    \includegraphics[width=\textwidth]{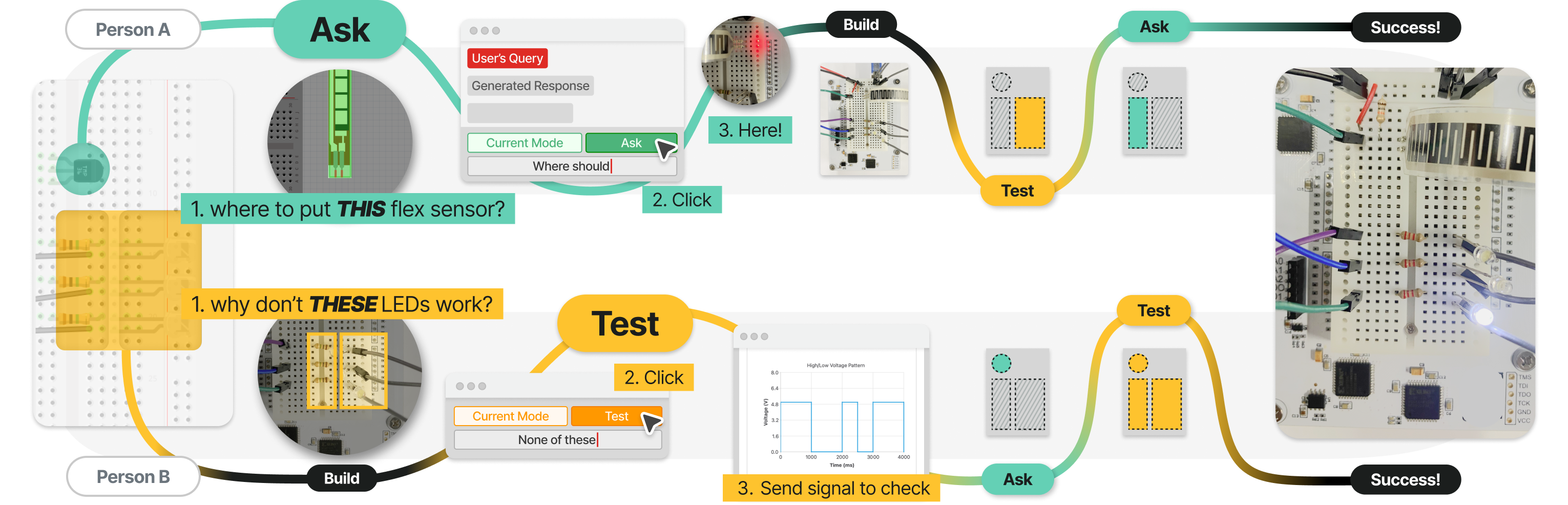}
    \caption{Two distinct personalized circuit-building workflows enabled by \systemName. Person A employs context-aware guidance by asking where to place a flex sensor, and receives \sout{component }highlighting assistance on the augmented breadboard. Person B follows a build-first approach, constructing the circuit from a schematic, asking "why don't these LEDs work?" and receiving automated test generation\sout{ capabilities}. Both workflows demonstrate successful circuit completion through individualized interaction strategies.}
    \Description{Two horizontal workflow tracks converging to shared "Success!" state. Top track (Person A): Grey "Person A" label, green "Ask" oval with breadboard inset showing flex sensor highlighted by green overlay and speech bubble query, chat interface with cursor on "Ask" button, black "Build" oval with glowing LED inset and breadboard with connected flex sensor, yellow "Test" oval, and "Success!" oval. Bottom track (Person B): Grey "Person B" label, black "Build" oval with breadboard inset showing LEDs/resistors highlighted by yellow overlay and speech bubble query, yellow "Test" oval with chat interface showing "Test" button and voltage pattern graph with square wave, and "Success!" oval. Both paths end at shared image of functional breadboard with flex sensor, illuminated LEDs, resistors, IC, and colored wires. Curved arrows connect stages, color-coded: green (Ask), yellow (Test), black (Build/Success).}
    \label{fig:Teaser}
\end{teaserfigure}


\maketitle

\section{Introduction}

The growing popularity of microcontroller platforms like Arduino\footnote{\url{https://www.arduino.cc/}} has enabled diverse end-user developers to create interactive electronic projects \cite{Bianchi_2023} and has contributed to the widespread adoption of \textit{physical computing} \cite{physicalcomputing}. Creators often utilize popular open-source tools, such as the authoring editor Fritzing \cite{Fritzing}, to easily design electronic diagrams and then assemble circuits on physical breadboards. In this process, step-by-step instructions gathered online on how to design, modify, prototype, and debug the desired circuit are crucial. Tutorials are particularly effective in helping break down complex circuit construction tasks into manageable steps, and this strategy has been adopted by numerous toolkits and research prototypes that scaffold circuit construction via interactive tutorials \cite{ElectroTutor, CircuitStyle, Ritschel_2023} and precompiled guided tests to verify the correctness of the final prototypes \cite{Muse, SensorViz, Heimdall}.

While these structured tutorials and tools are effective in breaking down complex circuit assembly tasks into manageable steps, this approach, however, conflicts with the practitioner's natural building processes and individual preferences \cite{DebugbyDesign, DesPortes_2019}. 
In the real world, creators employ highly idiosyncratic and bottom-up prototyping strategies, from selecting suitable circuits from online resources based on the availability of components to determining a specific order of implementation based on individual learning goals or iterative processes \cite{CrossedWires}. 
As such, while tutorials clearly scaffold users, they fail to leverage the creators' sense of agency and intuition and do not foster the development of problem-solving techniques or learning \cite{DebugbyDesign}.

This mismatch between prototyping tools and user behavior creates barriers that prevent effective exploration and debugging of physical computing projects. 
The root of this problem lies in the fragmented nature of existing support systems, which address specific aspects of physical computing development but fail to provide integrated solutions that adapt to individual circuit contexts. 
Tutorial-based systems \cite{HeyTeddy} and generative approaches \cite{TriggerActionCircuits} provide structured guidance, but they require makers to follow predetermined paths that may not align with the intended circuit topology. Circuit visualization tools \cite{CircuitSense, CurrentViz} excel at representing electrical states but operate independently from the construction process itself. Augmented breadboard systems \cite{BlinkBoard, SchemaBoard} demonstrate promising spatial guidance capabilities through LED-based instruction delivery, yet they remain disconnected from circuit design software and testing practices. Debugging tools \cite{Bifrost, Heimdall} bridge the monitoring between hardware and software to allow user-authored testing, yet cannot assist users in building personal applications with progress in hardware context integrated instructions.
In sum, existing solutions fall into two categories: \sout{they either }offer guidance through structured tutorials that need to be pre-authored by an instructor, enforcing a specific way to prototype and test a circuit~\citep[e.g.,][]{HeyTeddy, ElectroTutor}, or provide specific solutions or tools for the design~\citep[e.g.,][]{TriggerActionCircuits, Lin_2019}, construction~\citep[e.g.,][]{SchemaBoard, CircuitStack}, or debugging~\citep[e.g.,][]{Toastboard, CircuitSense} of electronic circuits, leaving the burden of deciding what to do and which strategy to adopt for different situations completely \sout{up }to the end developer.

To address this mismatch and bridge the gap, we present \systemName, an integrated development environment that enables personalized circuit prototyping workflow through hardware-contextualized guidance and in-situ testing capabilities, effectively linking the hardware with the digital representation of the system (Figure \ref{fig:Teaser}). Unlike fragmented existing tools, \systemName provides unified support by bridging physical computing design, construction, and validation processes. The system delivers adaptive guidance through a conversational agent that maintains real-time awareness of both visual circuit schematics and physical configurations. This dual awareness enables \systemName to provide circuit-specific visual cues directly on an augmented breadboard through real-time hardware sensing and automated testing capabilities. We demonstrate the design, implementation, and evaluation of \systemName through (1) a literature review and formative study ($N=12$) that informed the system's design goals and technical requirements, (2) the implementation details of our integrated prototyping platform, and (3) a usability study ($N=12$) demonstrating how hardware-contextualized guidance and validation enhance personalized circuit prototyping.
\section{Related Works} 
Building physical computing systems requires makers to navigate complex relationships between code, schematics, and physical hardware while following an often non-linear and exploratory process. We survey prior work across three dimensions that shape the physical computing experience: maker workflows, hardware-aware instructions, and circuit validation.

\subsection{Maker Workflows} 
Empirical studies consistently demonstrate that makers naturally employ highly personalized and exploratory prototyping strategies that promote a growth mindset and effective learning \cite{DebugbyDesign, CrossedWires, DesPortes_2019}. 
Theoretical frameworks like "reflection-in-action" describe design as a "thinking-by-doing" activity \cite{schön1983reflective}, emphasizing a conversational relationship between designer and medium where concrete prototypes lead to unexpected realizations \cite{Hartmann_2006, King_2017}.
These learning principles, where knowledge is built through hands-on tangible construction, \sout{was}\jw{were} proven effective for supporting diverse learners in physical computing education \cite{physicalCSE}.

However, this natural approach often conflicts with existing tools that impose structured, predetermined paths.
Makers face substantial challenges in physical computing, with most fatal faults due to incorrect circuit construction that are often misdiagnosed as software bugs \cite{Alessandrini_2022}.
Current educational approaches fail to provide comprehensive frameworks for identifying context-specific sources of bugs, highlighting systems thinking perspectives that support diverse debugging pathways \cite{McLaughlin_2025, DeLiema_2024}.
The complexity stems from needing to understand coding, electronics, and their interconnections, whereas current tools rely on error-prone manual processes that create barriers to scaling beyond prototypes \cite{MakeDevice}.
Research reveals that creative engineering work occurs during system architecture, yet tools operate at lower abstraction levels, creating tedious work and calling for more adaptive approaches \cite{Lin_2019, Bianchi_2023}.

While existing research advocates for personalized workflows and documents these challenges, available tools remain fragmented and misaligned with natural exploratory processes.
\systemName directly addresses this by enabling personalized circuit prototyping through unified support that bridges design, construction, and validation while enhancing makers' exploratory nature.

\subsection{Hardware-aware Instructions} 
Hardware guidance has evolved from structured to more adaptive, context-aware systems.
While effective at breaking down complex tasks, tutorial-based methods \cite{HeyTeddy, ElectroTutor} require predetermined paths that may not match individual circuit topologies or preferences.
Research into how-to videos shows the value of rich, contextual information for effective learning beyond mere instructions \cite{Yang_2023, MixT}.
Augmented breadboard systems offer spatial guidance through LED matrices that visualize component placement and electrical connections \cite{BlinkBoard, SchemaBoard, VisibleBreadboard}, representing a shift toward hardware-contextualized guidance with immediate visual feedback.
Hardware redesigns \cite{BitBlox} address cognitive load by bringing visibility to underlying breadboard connections, while CircuitStack \cite{CircuitStack} supports rapid and iterative circuit evolution.
Intelligent assistance tools include AutoFritz \cite{AutoFritz} for circuit autocomplete, Trigger-Action-Circuits \cite{TriggerActionCircuits} for generative design from behavioral descriptions, and CircuitStyle \cite{CircuitStyle} for reinforcing construction best practices.

Advanced approaches focus on bridging abstraction levels and offering novel interaction paradigms through different modalities.
Conversational agents like FritzBot \cite{FritzBot} take natural language descriptions from novice users and dynamically generate corresponding Arduino circuits and code, addressing component selection challenges.
\citet{Yusuf_2023} automates Arduino programming by generating hardware configurations and API usage patterns from natural language queries.
Visual programming environments such as Flowboard \cite{Flowboard} introduce flow-based programming that is conceptually closer to circuit diagrams than imperative code, providing immediate feedback that better reflects circuit behavior.
High-level design tools \cite{PolymorphicBlocks} and Hardware Description Language-based editors \cite{Lin_2021} allow designers to work while maintaining synchronization with visual representations.
Hybrid approaches, like VirtualComponent \cite{VirtualComponent}, assist with component value tuning through augmented reality (AR) overlays on physical breadboards, and VirtualWire \cite{VirtualWire} enables programmatic reconfiguration of breadboard connections.
SpatIO \cite{SpatIO} uses XR to support spatial component placement with virtual-to-physical transitions, while Proxino \cite{Proxino} blends virtual and physical circuit elements to facilitate distributed prototyping collaboration.
Finally, sensor-specific tools like SensorViz \cite{SensorViz} provide visualization across prototyping stages, from datasheet specifications to AR-based environmental interaction.

In summary, while there are rich hardware-aware systems and tools, they often require predetermined paths and operate in isolation from the real-time physical build context. 
\systemName offers a distinct advantage by providing adaptive guidance through a conversational agent that maintains \sout{real-time }awareness of both visual circuit schematics and physical configurations. 
This dual understanding enables to deliver contextualized support and circuit-specific visual cues directly on an augmented breadboard in a way that is integrated and responsive to the maker's actual physical progress.

\subsection{In-Situ Circuit Validation} 

Physical computing debugging is complex as bugs can reside in software, hardware, or their intersection, with novices often misdiagnosing hardware issues as software problems \cite{CrossedWires}.
Circuit visualization tools \cite{CurrentViz, CircuitSense} provide real-time flow visualization and automatic component recognition, but operate independently from construction workflows.

Integrated debugging environments bridge hardware-software gaps through systems like Bifröst \cite{Bifrost} for linked visualizations, Heimdall \cite{Heimdall} for remote inspection, and Toastboard \cite{Toastboard} for ubiquitous instrumentation.
Test-driven approaches like ElectroTutor \cite{ElectroTutor} extend software engineering concepts to hardware, but often require tests to be pre-designed by an instructor, which cannot be universally applied to any circuit. HeyTeddy \cite{HeyTeddy} automatically suggests users complete tasks throughout a tutorial, but also requires these tutorials and tests to be pre-specified.
Many solutions necessitate external instrumentation or separate monitors, removing information from the immediate code context.
Recent developments address these limitations through in-context approaches like Inline \cite{Inline} for direct code editor visualization, WiFröst \cite{WiFrost} for networked system instrumentation, and conversational debugging \cite{Bajpai_2024} for guided localization through natural language interaction.

Many debugging visualizations remain decoupled from the immediate construction context, requiring users to mentally map between separate interfaces and their physical circuits, or rely on complex setups or pre-existing exercises and tests compiled by an instructor.
\systemName addresses these limitations by providing a unified debugging experience within the circuit prototyping procedure.
Our approach provides automated testing tailored specifically to individual hardware through real-time input and output reads without additional setup or pre-made content.

\section{Formative Study}

We conducted a formative study to explore makers' challenges when developing and debugging physical computing systems with embedded platforms such as Arduino.
We recruited twelve participants (7 female, 5 male) via our institution's online community posting, with an average age of 23.75 \(\pm\) 3.05 years (range 19-29 years). 
We report means with standard deviations as M \(\pm\) SD.
They had diverse backgrounds in electrical engineering ($n=2$), computer science ($n=2$), mechanical engineering ($n=2$), industrial design engineering ($n=5$), and an exploratory major ($n=1$). 
Participants reported an average of 2.29 \(\pm\) 3.47 years of experience with embedded systems (range $<$1 to 13 years), see Table \ref{tab:formative_participants} for details.

\begin{table}[ht]
\footnotesize
\centering
\caption{Participant demographics and prior experience}
\label{tab:formative_participants}

\setlength{\tabcolsep}{3pt} 

\begin{tabular*}{\linewidth}{
l@{\hspace{4pt}}
c@{\hspace{4pt}}
c@{\hspace{8pt}}
@{\extracolsep{\fill}}
l c l c
}
\toprule
\textbf{ID} & \textbf{Gender} & \textbf{Age} & \textbf{Major} &
\makecell{\textbf{Coding}\\\textbf{Exp.}} &
\textbf{Program Lang.} &
\makecell{\textbf{Phys. Comp.}\\\textbf{Exp.}} \\
\midrule
P1  & M & 29 & EE & 13 yrs   & SystemVerilog & 13 yrs \\
P2  & M & 21 & CS/Math & 3--4 yrs & C             & $<$1 yr \\
P3  & F & 26 & EE & 3--4 yrs & Verilog       & $<$1 yr \\
P4  & F & 19 & Exploratory & 1--2 yrs & Python        & $<$1 yr \\
P5  & F & 20 & CS & 4--5 yrs & Python        & 1--2 yrs \\
P6  & F & 21 & ME & 1--2 yrs & Python        & $<$1 yr \\
P7  & F & 25 & ID & $<$1 yr  & Python        & $<$1 yr \\
P8  & F & 23 & ID & 1--2 yrs & Python        & 2--3 yrs \\
P9  & M & 27 & ME & 2--3 yrs & C++           & 2--3 yrs \\
P10 & F & 24 & ID & 2--3 yrs & Python        & 1--2 yrs \\
P11 & M & 26 & ID & 1--2 yrs & C\#           & 1--2 yrs \\
P12 & M & 24 & ID & 3--4 yrs & JavaScript    & 2--3 yrs \\
\bottomrule
\end{tabular*}
\Description{Demographics table for 12 formative study participants showing diverse backgrounds and experience levels. Participants aged 19-29, with a gender distribution of 5 males and 7 females. Academic majors include Electrical Engineering, Computer Science, Industrial Design, Mechanical Engineering, Mathematics, and Exploratory Studies. Coding experience ranges from less than 1 year to 13 years, with most participants having 1-4 years of experience. Programming languages include Python (5 participants), C variants, Verilog/SystemVerilog, C#, and JavaScript. Embedded systems experience is predominantly beginner level, with 7 participants having less than 1 year of experience and 5 participants having 1-3 years of experience.}
\end{table}

\textbf{Method}
Each study session lasted approximately 80 minutes and comprised three parts: introduction and consent, a demographics survey (10 minutes), a circuit-building task (50 minutes), and a semi-structured interview (20 minutes). 
Participants received compensation equivalent to \$15 USD in local currency.

\textbf{Circuit Prototyping Task}
We adapted \sout{the }\textit{Project 4: Color Mixing Lamp} from the Arduino Projects Book \cite{fitzgerald2012arduino} as it was rated a beginner-to-intermediate\sout{ project}, recommended for 45 minutes.
The task required \sout{participants }reading brightness values from three photoresistors and \sout{use them to }control an RGB LED's red, green, and blue channels.  
We allotted 50 minutes with a 5-minute grace period and provided an official schematic from the book. 
Participants used the internet for information search and debugging, but were restricted from using large language models beyond Google's default AI-summarized results.

\textbf{Task Result}
Four of twelve participants successfully built and coded the circuit. The participants who succeeded took an average time of 42'00" \(\pm\) 6'47" minutes. The reasons for failure consist of 1) misidentifying the type of the LED (common cathode to common anode), 2) wrong resistance value to construct the photoresistor voltage divider, and not accommodating for it in the code, 3) misconstructing the voltage divider for the photoresistor, 4) code discrepancy (wrong functions, not declaring pin mode). 

\textbf{Semi-Structured Interview}
The following debrief examined participants' information practices during the task, focusing on (1) tutorial search strategies, (2) format preferences and switching, (3) trustworthiness and accuracy judgments, (4) strategies for keeping track of multiple information sources, and (5) approaches to verification, error recovery, and reflection (Appendix \ref{appendix:formative}).
We recorded \sout{participants' }physical circuit-building processes during the task, screen-captured\sout{ their} programming and information searches, and voice-recorded the interview sessions.
After transcribing and translating the interviews into English, we applied open, axial, and selective coding for qualitative analysis \cite{GroundedTheory}.  
As a result, we introduce three key challenges identified across participants during physical computing projects, particularly highlighting the mismatch between existing tool design and makers' natural building processes.

\subsection{Challenge 1: Distinct Building and Information Retrieval Strategies}

Participants showed idiosyncratic strategies in interactive circuit prototyping, including starting component choices and task segmentation\sout{---}\jw{, }that is to say, there was no single pattern shared across all participants. Some participants broke down complex tasks into smaller, manageable segments that reflected their individual problem-solving approaches (P1-P2, P9-P12). For example, P1 described breaking down the task into reading the photoresistors' brightness value first, setting up the LED, and later merging them. P2, P9, P11, and P12 echoed this incremental segmentation strategy; P9 referred to \textit{"dividing the task into some fundamental elements,"} while P12 used the metaphor of doing \textit{"baby steps, little achievements along the way."} P10 separated the task into component types and tried using a single photoresistor first, then added the remaining two photoresistors. 
In stark contrast, other participants (e.g., P4, P6-P7) chose to build everything at once without breaking down the circuit into smaller parts or tasks. They preferred to build the entire circuit and test it at once at the end. This was not a particularly effective strategy, as acknowledged by P4: \textit{"I think that was a problem... I have to basically fix all of the parts at once."} However, P4 recognized the value of component-level testing for more complex projects, explaining: \textit{"For [what I'm currently working on]... just at first try to test one motor and one wire and then if that is right, then I use the same value to calculate the whole thing."}

When we asked how they previously selected relevant tutorials and information for the circuit, we also received descriptions of numerous and diverse approaches. Some participants prioritized images (P1-P2, P6, P8-P10, P12), while others preferred video tutorials because they supported \textit{"understanding the whole process"} (P4) but also because they could pause them anytime if they wanted to screen capture images of circuits or schematic diagrams (P3, P5-P6, P10, P12). Another notable difference among participants was the choice of preferred language, as not all participants were native English speakers. As such, people (P4, P7-P8, P11) explained feeling more comfortable using their native language for information retrieval. P4 explained their keyword selection strategy: \textit{"I chose my keyword based on my experience because I'm not familiar with English, so I learned in school how this [photoresistor] is called in [my mother tongue]."} There was also the issue of domain-specific languages (i.e., the technical jargon or the name of specific hardware components), as not every participant shared the same engineering knowledge. Participants referred to elements in this circuit using pronouns or determiners such as "it," "this," and "that" (P3, P5, P7, P10-P12).
P10 highlighted the communication challenges when searching online: \textit{"It's hard to communicate what this is in texts on the Internet, saying 'What is this box with two arrows?', or 'What does that mean?' They won't understand what that is."} 

\textbf{Takeaway}: Our interviews revealed that creators adopt very different strategies for constructing circuits or searching for relevant information online, and that they all value different types of information (images or videos) or choose vocabulary that reflects their prior knowledge, preferences, and expertise.

\subsection{Challenge 2: Mismatched Between Circuit and Tutorial Diagrams}

All participants discussed how differences between the visual representation of a circuit and the circuit diagram, or even \sout{when}\jw{between} components on their breadboard \sout{looked different from}\jw{and} those in the tutorials, created confusion and misunderstanding.
This separation between equivalent representations (i.e., digital vs. physical) was already studied in the literature \cite{SchemaBoard} and was further corroborated in our findings. For example, nine participants reported struggling to identify components and pins (P1-P4, P7-P10, P12) when tutorial visuals did not match their prior knowledge or the components' physical appearance.
P1 experienced wiring errors due to incorrect assumptions about LED pin ordering, initially expecting "ground, red, green, blue" but finding the actual arrangement was "red, ground, green, blue."
Two participants (P6, P7) expressed frustration with the way power connections (VCC and GND pins) were rerouted directly to the components without passing through the breadboard, or how different components should have been grounded to the same node of the net.
Similarly, P3 made a \textit{"huge mistake"} by misidentifying photoresistors as a diode in the provided schematic, stating, \textit{"I was familiar with arrow symbols for diode,"} and only realizing the error after watching a clarifying video.
P12 observed that the diagram in \sout{their }chosen tutorial had pins arranged differently from what they had built based on general knowledge, leading them to tear down and rebuild the circuit to match the tutorial exactly.

Another source of frustration was identifying components and their intended usage. P2 observed that different online sources often presented conflicting visual representations, \textit{"the documentation... was different from what you're expecting from the components that you have""}. Several participants (P2, P8, P10-P11) struggled with tutorial applicability when similar components were used for different purposes, \textit{"two tutorials look similar but were made for different purposes"}--P8. As a solution, users turned to the text descriptions in the tutorial to get a better grasp of the circuit topology and its intended usage (P4, P6-P11), leaving some to conclude (P10, P11) that constant adjustment and modification are needed to extract relevant information from tutorials. 

\textbf{Takeaway}: All the above comments highlight the frustration of creators, especially those with little engineering expertise, in forming a clear mental model that bridges the diagram representation of a circuit and its physical instantiation in the hardware. Any mismatch in terms of visual representation between the two, in terms of circuit structure or components, creates uncertainties, confusion, or simply slows down the process and confidence with which the circuit can be assembled.

\subsection{Challenge 3: Manual and Iterative Circuit Debugging}
Circuit validation emerged as the most significant barrier to successful physical computing prototyping. Multiple participants (P5, P6, P7, P8) lacked systematic testing knowledge and resorted to guesswork rather than methodical verification approaches. P6 explicitly acknowledged this limitation: \textit{"I think it's hard to do [verify]. So it's more of a guess that it will work"}, while P7 admitted forgetting fundamental testing procedures: \textit{"But I forgot that there is that kind of [testing individual components before building the whole thing] process."} This finding aligns with prior work documenting debugging challenges in physical computing education, where circuit construction errors often compound software bugs, creating complex failure scenarios that novices struggle to diagnose systematically.

Existing testing approaches required external tools that operated independently from circuit design contexts, contrasting sharply with participants' requests for integrated solutions. 
While participants relied on disconnected debugging methods--serial monitor observations (P1-P2, P4-P5, P9-P11), manual component commanding through temporary code modifications (P2, P9), or external measurement tools like multimeters (P1)--six participants (P1-P3, P5, P10, P12) explicitly desired automated validation systems integrated directly into circuit design workflows. 
P9 exemplified current fragmented approaches: \textit{"turn[ing] each LED one by one directly through the code without anything connected... turning each (LED) 'high' one by one"} to verify connections, while P10 conceptualized integrated alternatives: \textit{"a LEGO-style tutorial with step-by-step checkpoints that would show the simulated values of a circuit that should be expected."} 
However, even general-purpose AI assistance proved inadequate, as P10 experienced: \textit{"There were times when I followed everything that the [AI] was telling me, and it still didn't work... they kept telling me very general answers that I had already checked."} 
P5's request for immediate contextual feedback—\textit{"if the program could tell if I'm wiring the components wrong"}, alongside P1's advocacy for \textit{"built-in instrumentation"} with \textit{"measurement ports"}, highlights the need for circuit-aware debugging support that provides specific, actionable guidance rather than requiring manual correlation of disparate information sources or generic troubleshooting advice.

\textbf{Takeaway}: Overall, users expressed frustration in identifying the source of errors and showed interest in methods that would be aware of the circuit under test, suggest strategies to narrow down the possible cause of errors, and guide them through strategies to measure the expected output.
\begin{figure*}[h!]
    \centering
    \includegraphics[width=0.9\linewidth]{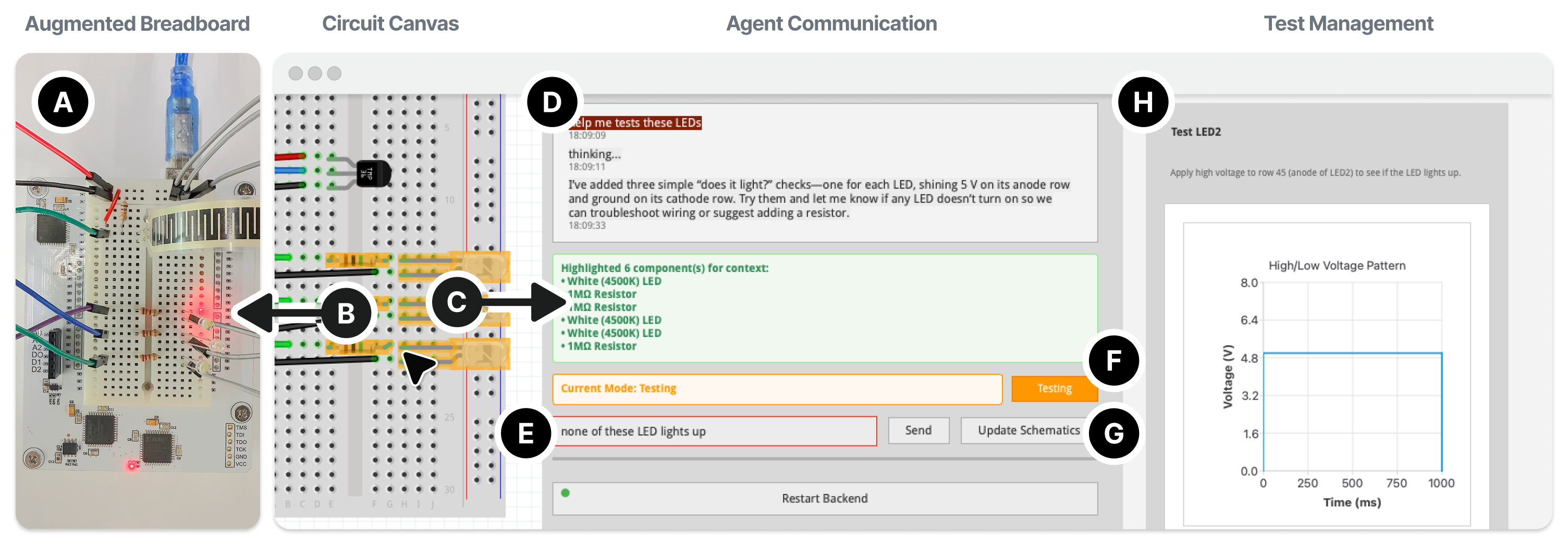}
    \caption{Overview of \systemName interface architecture: (A) an augmented breadboard with \sout{embedded }LED row indicators for physical guidance, (B) a circuit design canvas enabling circuit configuration, visualization, and \sout{component selection to highlight on the }breadboard \jw{highlighting}, (C) component selection for conversational context reference, and (D) an AI agent communication\sout{ interface}. Users interact through (E) natural-language text input to query circuit configurations and components. The interface provides (F) mode switching between \textit{Ask} and \textit{Test}\sout{ operations}, (G) schematic synchronization to update circuit context, and (H) ad-hoc test\sout{ management} for systematic hardware verification.}
    \label{fig:system-overview}
    \Description{System interface showing an augmented breadboard on the left with LED-lit rows, a central circuit design canvas displaying electronic components, a conversational AI chat interface on the right, and a test management panel with voltage measurement graphs and control buttons.}
\end{figure*}

\section{\systemName System}

\systemName is a circuit design and construction support tool that makes the realization of a circuit design easier by connecting software visualization with physical circuits and integrating expressive agent chat queries with context inclusion (Figure \ref{fig:system-overview}).
The system was structured in response to the circuit realization challenge noted during the literature review and the formative study. 
\systemName was designed with three goals: 1) \textbf{let the user ask questions about their own circuit and components} without having to conform to any preprocessed tutorial; 2) the user can receive \textbf{guidance on where to physically wire and/or apply tests directly in the hardware}; and, 3) \textbf{the system can suggest and perform hardware tests} that help the user find faults in a circuit by probing it (see Appendix \ref{appendix:system_interface} for details).

\subsection{Design Goal 1: Supporting Idiosyncratic Prototyping} 

\begin{figure*}[h!]
    \centering
    \includegraphics[width=0.9\linewidth]{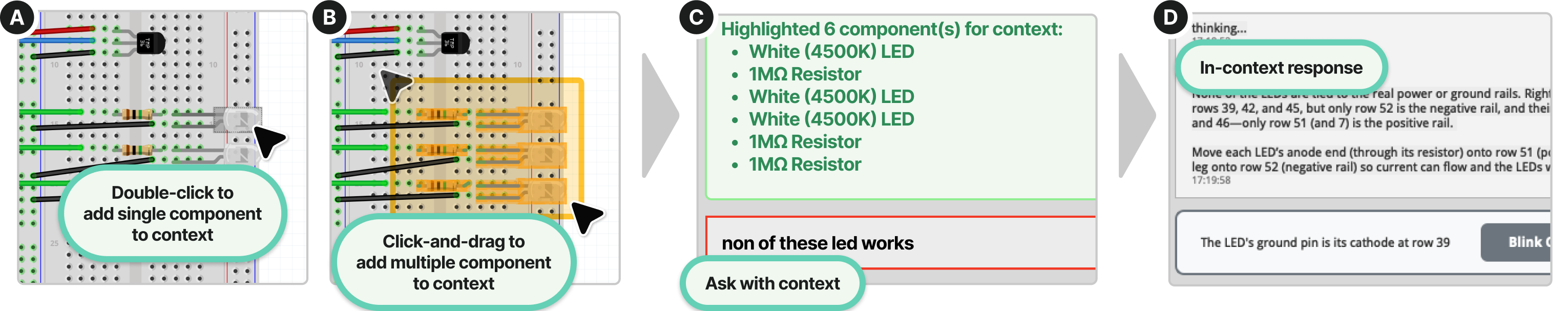}
    \caption{Context-aware component selection workflow for conversational interaction. Users can add circuit components to the\sout{ AI's} conversational context through (A) double-click\sout{ing} individual components in the circuit design interface, or (B) click-and-drag\sout{ selection} for multiple components. (C) With highlighted components as \jw{added} context, users can query using natural language and pronoun references. (D) The system parses the contextual component information to provide targeted guidance and responses.} 
    \label{fig:context-selection}
    \Description{Four-panel workflow showing component selection methods: Panel A shows a single LED being double-clicked, Panel B shows multiple components being selected via click-and-drag, Panel C displays a context panel listing selected components with natural language query input, and Panel D shows the AI agent's contextual response about LED connections.}
\end{figure*}

To better scaffold makers in their own circuit modifications and explorations, \systemName provides personalized guidance that adapts to the current circuit configuration at any arbitrary starting point. 
By integrating a written query pipeline with included circuit context into the system interface (Figure \ref{fig:context-selection}), the user can have an interactive conversation about the construction strategy and process for this specific circuit, or the characteristics (e.g., pinout) of a specific component. 
With the circuit's context, the user can ask general questions about the circuit configuration (e.g., \textit{Is the circuit correct? Are there any misconnections?}), and by visually selecting specific components or a portion of the circuit on the interface (Figure \ref{fig:context-selection}A, B), direct questions about the highlighted parts alone. 
Context-awareness also allows users to refer to components using generic pronouns and terms such as "this," "here," and "these" in \sout{their }queries (Figure \ref{fig:context-selection}C and \ref{fig:system-overview}C). For example, the user can ask \textit{What is this component?} or \textit{How shall I connect this component?}, or \textit{Where is the ground pin of this LED?} and the system can produce an informed response using as context the current circuit topology and the specified subset of components highlighted through the system interface (Figure \ref{fig:context-selection}D).

These mechanisms address Challenges 1 and 2 from the formative study by providing ways for individual builders to have different execution strategies, formed through a dialogue with the agent, for any arbitrary tutorial that serves as a starting point.

\subsection{Design Goal 2: Guiding with Hardware-Context Awarness}

\begin{figure}[h!]
    \centering
    \includegraphics[width=0.74\linewidth]{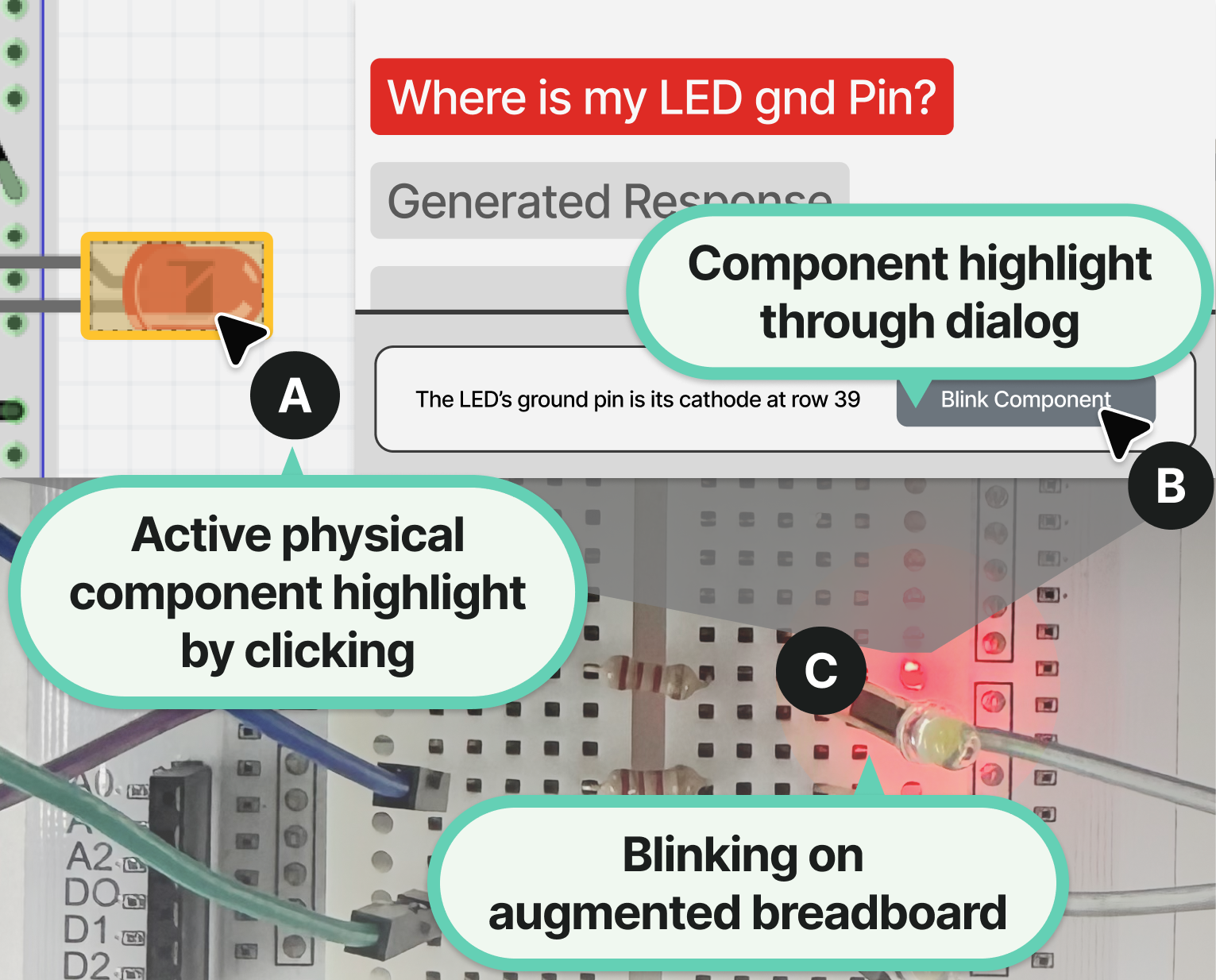}
    \caption{Two methods for providing in-situ \jw{breadboard} guidance\sout{ through an augmented breadboard}: (A)\sout{ Active component highlighting:} users click circuit components \sout{in the design interface }to illuminate corresponding breadboard rows. (B)\sout{ AI-driven guidance:} the conversational agent automatically activates (C) LED indicators and provides interface buttons to repeat the highlighting sequence.}
    \label{fig:walkthrough-guidance-led}
    \Description{Split-panel interface showing LED guidance methods: Top panel displays circuit design software with component selection triggering LED highlights, bottom panel shows physical augmented breadboard with illuminated rows and AI chat interface containing guidance buttons for component location assistance.}
\end{figure}

\begin{figure*}[h!]
    \centering
    \includegraphics[width=0.75\linewidth]{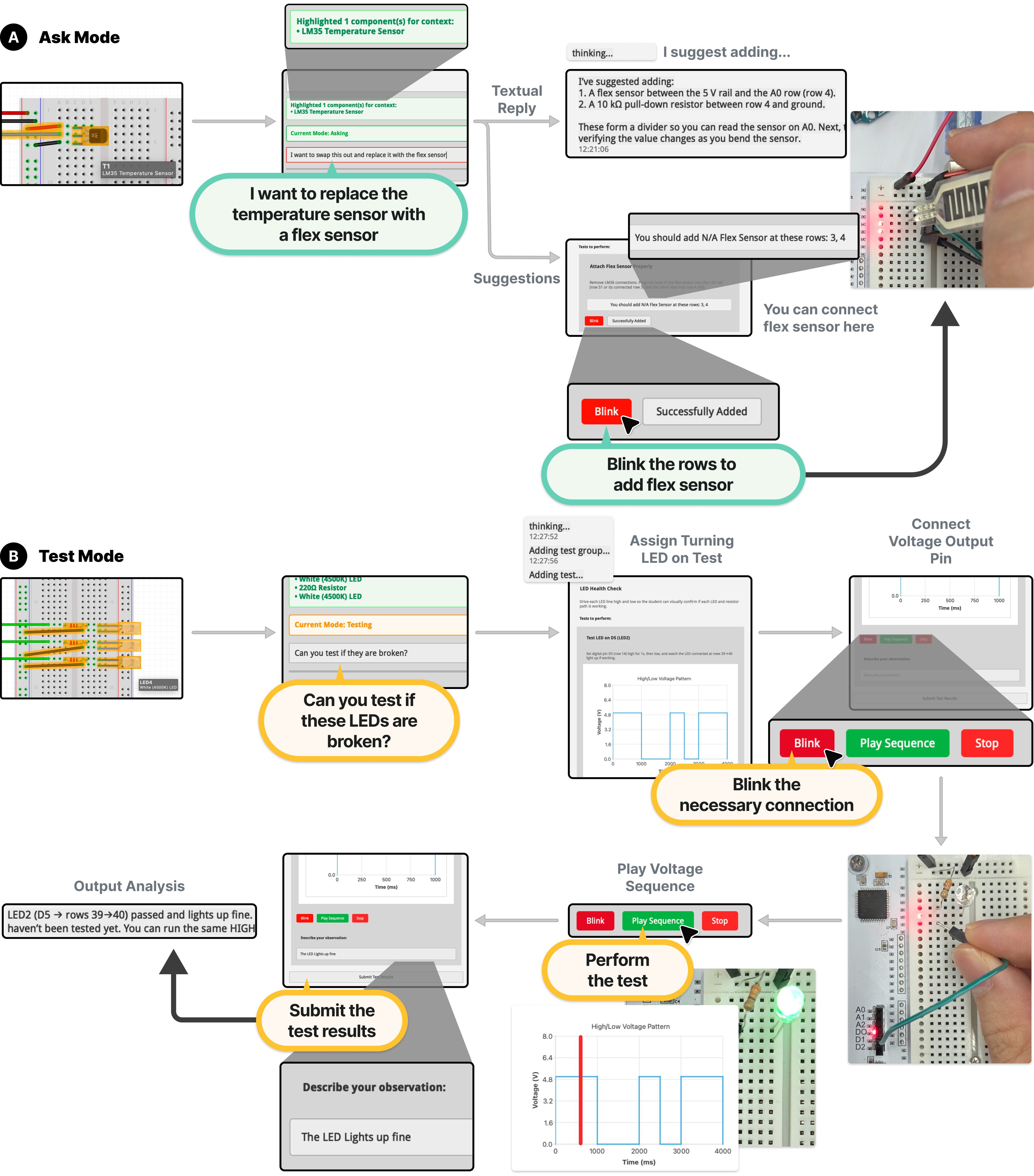}
    \caption{Interaction flow of \systemName. (A) \sout{During }\textit{Ask Mode}: users can select components \jw{in}\sout{to} context and submit a query. The system responds \jw{with either}\sout{either with} textual guidance (no physical action required) or suggestions \jw{highlighting}\sout{that highlight} where to modify the circuit (Figure \ref{fig:test-types}B). (B) \sout{In }\textit{Test Mode}: users \jw{can }select \jw{which }components to \sout{be}test\sout{ed}. The system generates test procedures, groups related tests, and may require probe connections (Figure \ref{fig:test-types}A). Users can again highlight required I/O pins before running a test sequence. \jw{Users can submit results to receive processed feedback or interpretation.}\sout{After observing component behavior, users submit results, and the system provides basic feedback or interpretation.}}
    \label{fig:test-and-ask-flow}
    \Description{Two-section workflow diagram showing Ask Mode and Test Mode interactions. Ask Mode: Breadboard with LM35 sensor and LEDs, component selection overlay with cursor, text interface showing highlighted component and user query about replacing with flex sensor, system suggestions panel with "Blink" button, and physical flex sensor being connected. Test Mode: Breadboard with orange selection box around multiple components, text interface with highlighted components and test query, test management panel with voltage pattern graph and control buttons, physical breadboard with glowing LED, and output analysis panel with test results and submission interface. Arrows and speech bubbles connect workflow steps.}
\end{figure*}

\systemName supports in-context physical hardware guidance by utilizing an open-sourced augmented breadboard with a built-in LED row indicator. 
\systemName can make an explicit row indication to the user by blinking the LED embedded in rows that it needs to make reference to (Figure \ref{fig:walkthrough-guidance-led}C). The system supports this position visualization in two ways (Figure \ref{fig:walkthrough-guidance-led}): 
\textbf{1) Active Mode} The user \sout{can click}\jw{clicks} on a specific component on the GUI breadboard and see the rows on the augmented breadboard lit up corresponding to the location where the selected component should be (Figure \ref{fig:walkthrough-guidance-led}A), similar to the previous works \cite{SchemaBoard}.
\textbf{2) Dialog Mode} \sout{The other}\jw{Another} option is through dialogue with the agent, where the user \sout{might }asks the agent to highlight a specific pinout of an IC or simply highlight the connections that the user should make. Then the system will display a GUI button under the chatbox, where the user can click to light up on the physical breadboard's rows of interest (Figure \ref{fig:walkthrough-guidance-led}B). For example, the user might select a 555 IC and ask the system to display where the "discharge" pinout \sout{of the said IC }is on the physical breadboard.

These interactions address Challenge 2 from the formative study by providing a one-to-one, easy-to-follow, explicit mapping for the user to realize the software design in physical hardware. And an explicit way that the system can make reference to the in-situ location of the component.

\subsection{Design Goal 3: Verifying Circuit via In-Situ Tests}

As a response to the formative study Challenge 3, \systemName allows users to identify circuit errors via a set of tests \sout{that are }dynamically generated to suit the current circuit, without requiring end developers to augment the code or instructors to prepare \sout{the }tests in advance. 
Specifically, \systemName helps users verify the current hardware configuration and, if \sout{the current configuration}\jw{it} contains a potential mistake, suggests a fix accordingly.
As a response to the formative study's Challenge 3, \systemName allows users to identify circuit errors via a set of tests \sout{that are }dynamically generated to suit the current circuit, without requiring end developers to augment the code or instructors to prepare the tests in advance. Specifically, \systemName helps users verify the current hardware configuration and, if \jw{it}\sout{the current configuration} contains a possible mistake, \jw{suggests}\sout{makes a suggestion} to fix \sout{it }accordingly. 
There are two \sout{operation }modes in which the system can operate: \textbf{1) \textit{Ask mode}}, the default mode where the system's goal is to provide information about the circuit and is limited to only making suggestions without assigning any tests. \sout{For example, }Figure~\ref{fig:test-and-ask-flow}A shows how user could inquire the system on how they should replace the temperature sensor as well as how the system would response with text and suggestions; or \textbf{2) \textit{Test mode}}, where the system's goal shifts to help users localize bugs by providing physical circuit tests that ask the user to perform an action and collect relevant data. For example, Figure~\ref{fig:test-and-ask-flow}B shows how the user can request the system to test one of the LEDs in the circuit. The system replies back with a simple voltage output test that the user can perform. The user can highlight where to connect output probes with the "blink" button.
After connecting all the required connections, the user then performs the test and sees how the LED responds to the voltage output graph. 
After seeing that the LED lights up fine, the user can submit the result, and the system replies \sout{back}with a result analysis. The user can choose the system operation mode via a toggle button (Figure \ref{fig:system-overview}F).

\begin{figure}[h!]
    \centering
    \includegraphics[width=\linewidth]{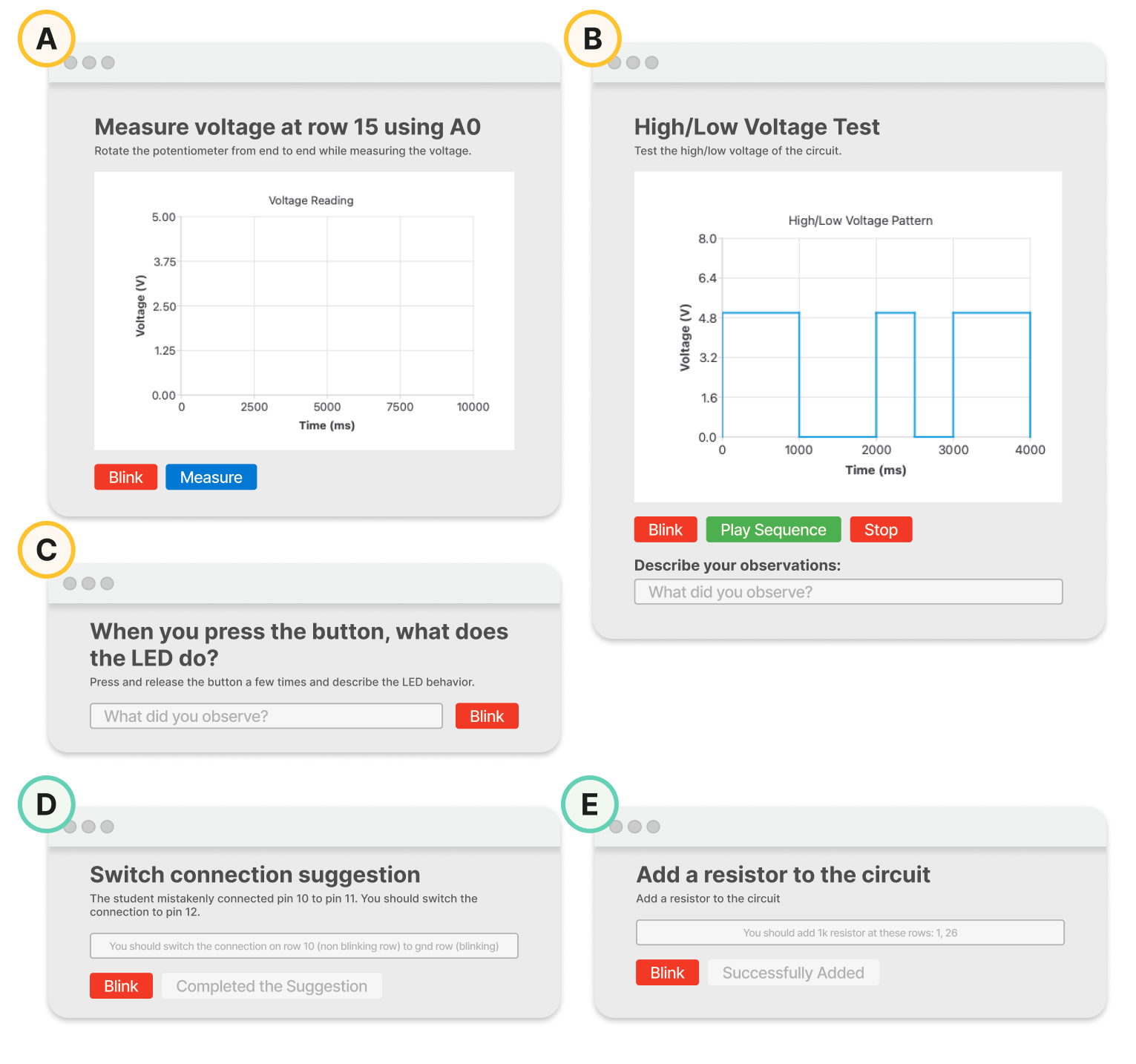}
    \caption{Five types of adaptive tests and suggestions generated by \systemName\sout{ in \textit{Test Mode} and \textit{Ask Mode}}. \sout{In }\textit{Test Mode}\sout{, the system} provides three diagnostic test types: (A) Voltage Measurement Test for probing circuit voltages at specific locations, (B) High/Low Voltage Pattern Test for analyzing the circuit response to digital signal behavior over time, and (C) Visual Inspection Test for observing component responses to user interactions. \sout{In }\textit{Ask Mode}\sout{, the system} generates: (D) Connection Suggestion for correcting miswired\sout{ connections}, and (E) Component Suggestion for adding missing circuit elements.} 
    \label{fig:test-types}
    \Description{Five interface panels arranged in two rows showing different test and suggestion types. Top row (yellow border): Panel A shows a voltage measurement interface with an empty line graph and "Blink/Measure" buttons, Panel B displays a high/low voltage pattern test with bar chart showing voltage spikes and "Blink/Play Sequence/Stop" controls, Panel C shows visual inspection test with text prompt about LED behavior and observation input field. Bottom row (green border): Panel D presents a connection suggestion interface with guidance text about switching pin connections and "Blink/Completed" buttons, Panel E shows a component suggestion interface with instructions to add a resistor and "Blink/Successfully Added" controls.}
\end{figure}

\subsubsection{Test and Suggestion Types}
The system provides 3 types of in situ tests and 2 types of suggestions that the conversation agent is capable of (Figure \ref{fig:test-types}). 
These tests and suggestions are generated on demand upon user requests or suggested by the system as a result of a previous question. 
The three tests consists of \textbf{1) Voltage Measurement Test} is assigned when the system wants to probe a part of the circuit for its voltage value (Figure \ref{fig:test-types}A) \textbf{2) Signal Output Test} is assigned when the system wants to see how a certain component respond to a digital high and low (Figure \ref{fig:test-types}B). \textbf{3) Visual Inspection Test} is assigned when the system wants the user to observe a certain behavior of a circuit like whether an LED respond to a button press (Figure \ref{fig:test-types}C); Voltage and Signal Output tests probe the existing circuit by generating a time-varying voltage signal through the augmented breadboard built-in Digital-to-Analog Converter (DAC). 
The two suggestions that the system can make are \textbf{1) Connection Suggestion}, which suggests the user to correct a possibly miswired connection at certain rows (Figure \ref{fig:test-types}D); \textbf{2) Component Suggestion}, which suggests the user to add a component to the identified place on the breadboard (Figure \ref{fig:test-types}E). 

\subsection{Implementation}

\begin{figure*}[h!]
    \centering
    \includegraphics[width=0.8\linewidth]{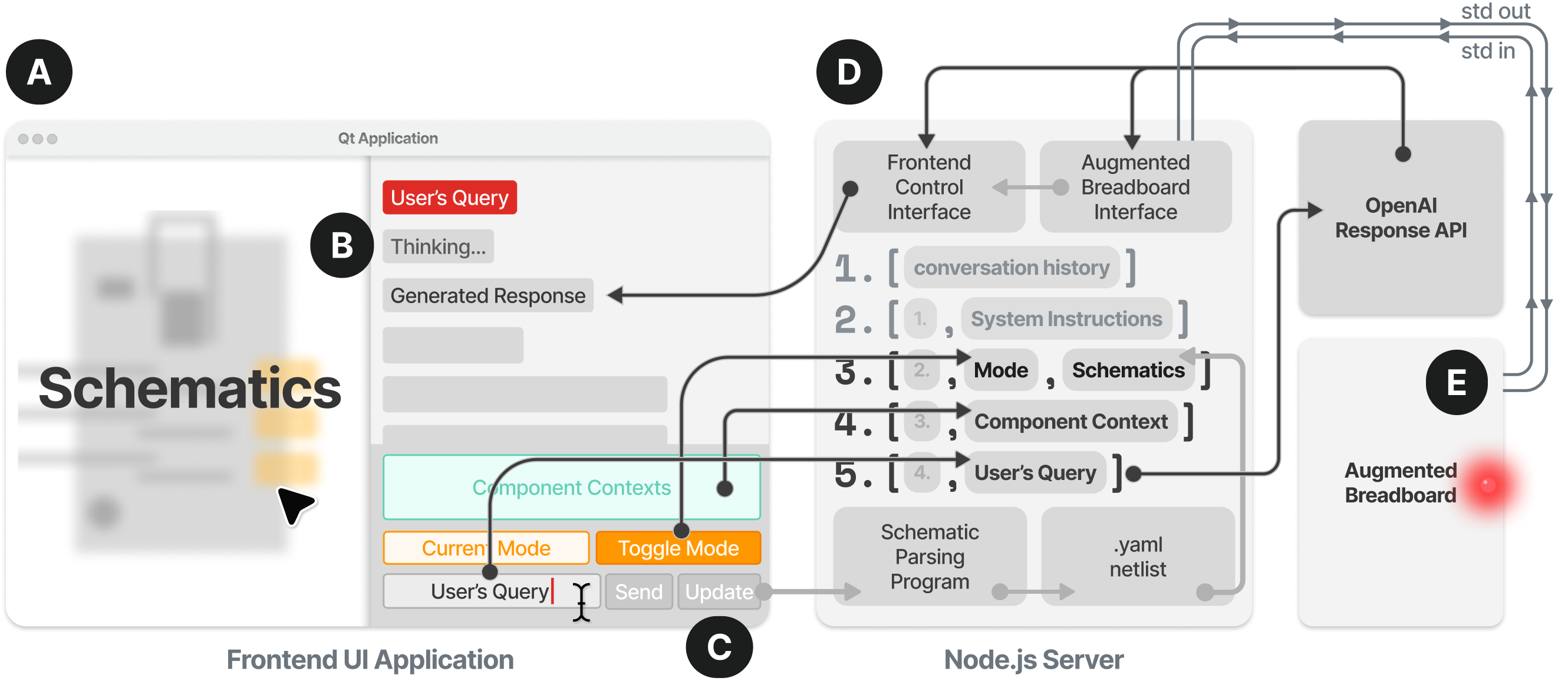}
    \caption{\systemName system architecture showing the \sout{complete }data flow and processing pipeline. (A) The frontend Fritzing extension captures user interactions and (B) displays system state while preparing answers or tests (C) and circuit designs, generating XML netlists that are converted to YAML format by the backend. (D) Node.js \sout{backend }server maintains conversation history, mode state, schematic context, and component selections while orchestrating AI queries through OpenAI API integration. (E) The augmented breadboard receives control commands and component highlighting instructions for real-time physical guidance\sout{ and hardware interaction}.} 
    \label{fig:architecture}
    \Description{Three-component architecture diagram with data flow arrows. Left: Qt Application window showing blurred schematics view, chat interface with "User's Query", "Thinking...", and "Generated Response" fields, control buttons for component contexts and mode selection, and mouse cursor pointing to Send button. Center: Node.js Server box displaying processing modules for user inputs, conversation history, system instructions, and component context, with connection to OpenAI Response API and schematic parsing program generating YAML netlist. Right: Augmented Breadboard hardware with red glowing indicator light. Arrows indicate bidirectional data flow between all components.}
\end{figure*}

\systemName integrates a modified Fritzing application \cite{Fritzing} with unmodified BlinkBoard \cite{BlinkBoard}; an open-sourced extended breadboard featuring row-indication LEDs and analog/digital I/O pins.
The BlinkBoard \cite{BlinkBoard} was chosen for simplicity in both user interaction and software implementation, controlled over a USB serial communication using JSON command format. The control commands consisted of: (1) executing row-indicator LED patterns (on, off, blinking, or blinking slowly); (2) outputting voltage or pulse-width modulation (PWM) from output pins; and (3) reading voltage inputs from the input pins.
The architecture comprises: (1) a user frontend application developed as a Fritzing extension (Figure \ref{fig:architecture}A); (2) a Node.js backend server to communicate with the OpenAI API and the hardware (Figure \ref{fig:architecture}D), and (3) a BlinkBoard that communicates with the server (Figure \ref{fig:architecture}E). The server effectively bridges between the Fritzing frontend (via stdin/stdout), the augmented hardware (via serial), and the OpenAI o4-mini model\footnote{\url{https://openai.com/api/}}. 

The Fritzing frontend extension (Figure \ref{fig:architecture}A) provides an interface for the user to 1) enter a query prompt; 2) select between 'ask' and 'test' mode; 3) add component contexts to the query prompt; 4) update the schematic configuration context; 5) interact with all the tests; and 6) control the augmented breadboard LED activity and I/O pins. All of which are sent to be processed in the backend. To include component context in the query, the frontend application extracts component metadata, including component identifiers, names, and pin connection mappings, and passes it to the server along with the subsequent query that the user sends. The frontend displays the response generation status like "thinking" or "adding tests" (Figure \ref{fig:architecture}B) to give visual latency feedback. The frontend was developed and implemented in C++, using the QT framework.

The backend (Figure \ref{fig:architecture}D) handles forwarding and structuring the user's query prompt to the OpenAI response API. Before sending the query, the server evaluates system state changes (e.g., mode changes, schematic updates) and incorporates relevant component context (if any) into the conversation array with the 'developer' role designation. User input messages are subsequently appended with the 'user' role designation. Initial system activation triggers the inclusion of comprehensive system instructions to establish operational parameters. The conversational agent extends its capabilities beyond textual responses through structured function calls that enable hardware control operations, including augmented breadboard LED activation sequences and schematic state queries. Function call implementations utilize predefined parameter schemas to ensure predictable agent behavior and maintain system reliability. Function call also allows the agent to generate context-aware tests for the user. Testing functions are exclusively available during \textit{Test mode} operation, preventing inadvertent activation during standard conversational interactions. Each test function incorporates hardware constraint specifications within its parameter schema (e.g., valid breadboard row ranges 1-50, timing parameters specified in milliseconds) to ensure safe and bounded operation. Upon receiving the XML netlist from the frontend as a result of the user updating the schematic context (Figure \ref{fig:architecture}C\sout{, Figure}\jw{ and} \ref{fig:system-overview}G), the server passes the netlist to a Rust-based binary parser that processes the XML, removes redundant information (e.g., duplicates and empty rows), and converts it to YAML format.

The breadboard can be controlled via serial communication and JSON commands (Figure \ref{fig:architecture}E). The server provides endpoints that accept augmented breadboard control parameters for the frontend to interface with. The conversational agent accesses identical functionality via structured function calls. After receiving a control request, the server generates the corresponding JSON and transmits it to the board via serial communication. The breadboard also supports voltage measurements through built-in ADC pins and digital signal output via configurable digital pins, with measurement results returned to the frontend application for display. The system achieves a latency of 8.23 \(\pm\) 0.69 ms for voltage output commands and 12.25 \(\pm\) 0.44 ms for analog read commands, measured across 500 repeated trials. Voltage values are addressed at millivolt resolution in the firmware, enabling theoretical precision to 1mV.

\section{User Study}
To evaluate how makers with diverse backgrounds adopt different physical prototyping practices, we conducted a usability study with \systemName. 
Our research question focused on understanding how users leverage hardware-contextualized guidance and in-situ testing capabilities during physical circuit prototyping, examining both individual workflow preferences and system effectiveness.
We designed a controlled study where participants worked with \sout{an}\jw{the} identical circuit containing deliberate bugs to observe and compare behavioral patterns and problem-solving approaches. Both quantitative metrics (completion times, error rates, feature usage, post-task survey) and qualitative insights through semi-structured interviews were collected to evaluate how the system supports personalized circuit construction workflows.

\newcolumntype{L}{>{\raggedright\arraybackslash}X} 
\newcolumntype{A}{>{\centering\arraybackslash}p{2.4em}} 
\newcolumntype{E}{>{\centering\arraybackslash}p{3.0em}} 
\newcolumntype{P}{>{\centering\arraybackslash}p{2.2em}} 

\begin{table*}[h!]
\footnotesize
\centering
\caption{Participant demographics and prior experience}
\label{tab:usability_participants}

\setlength{\tabcolsep}{3.5pt}
\renewcommand{\arraystretch}{1.05}

\begin{tabularx}{\linewidth}{
  l E A l   
  E L       
  E L P     
  L         
}
\toprule
\textbf{ID} & \textbf{Gender} & \textbf{Age} & \textbf{Major}
& \multicolumn{2}{c}{\textbf{Programming}}
& \multicolumn{3}{c}{\textbf{Physical Computing}}
& \textbf{AI Use} \\
\cmidrule(lr){5-6} \cmidrule(lr){7-9}
 &  &  &
& \textbf{Exp.} & \textbf{Lang.}
& \textbf{Exp.} & \textbf{Edu.} & \textbf{Prj.}
&  \\
\midrule
P1  & M & 26 & CE & 4--5y & C++ & 3--4y & Univ. course & 5+ & ChatGPT\textsuperscript{a, b} \\
P2  & M & 30 & CS & 4--5y & Go & 1--2y & Univ. course & 3 & Copilot\textsuperscript{c} \\
P3  & M & 29 & EE & 13y & C, SystemVerilog, Python & 13y & Univ. course & 5+ & Chatbot\textsuperscript{c} \\
P4  & M & 23 & Physics & 1--2y & Python & < 1y & High school & 1 & None \\
P5  & M & 27 & ID, HCI & 4--5y & Python & 4--5y & Workshop, Univ. & 5+ & ChatGPT, Gemini\textsuperscript{a, d} \\
P6  & F & 25 & ID & 3--4y & C & 2--3y & Univ. & 5+ & ChatGPT\textsuperscript{a} \\
P7  & M & 22 & ID & < 1y & Python & 1--2y & Univ. & 2 & ChatGPT\textsuperscript{d} \\
P8  & F & 25 & CT & 5--6y & Java & 5--6y & Self-taught & 5+ & ChatGPT\textsuperscript{a} \\
P9  & F & 24 & ID & 1--2y & Python & 1--2y & Univ. & 2 & ChatGPT\textsuperscript{a} \\
P10 & F & 25 & ID & 3--4y & C & 1--2y & Self-taught & 2 & ChatGPT\textsuperscript{a} \\
P11 & M & 20 & ID & 4--5y & Python & 4--5y & Univ., MOOC & 5+ & Claude, ChatGPT\textsuperscript{b} \\
P12 & F & 23 & ID & 1--2y & C, Python & 2--3y & Workshop, Univ. & 4 & ChatGPT\textsuperscript{b} \\
\bottomrule
\end{tabularx}
\Description{Demographics table for 12 usability study participants with comprehensive experience profiles. Participants aged 20-30 with a gender distribution of 8 males and 4 females. Academic backgrounds span Computer Engineering, Computer Science, Electrical Engineering, Physics, Industrial Design, HCI, and Communication Technology. Programming experience ranges from less than 1 year to 13 years, with most participants having 1-5 years of experience. Programming languages include Python (6 participants), C variants (4 participants), plus Java, Go, and SystemVerilog. Physical computing experience ranges from less than 1 year to 13 years, acquired through university courses (8 participants), workshops, self-teaching, and MOOCs. Project experience spans 1-5+ completed projects. AI tool usage is prevalent, with 11 of 12 participants using AI assistance, predominantly ChatGPT, with use cases including circuit debugging and troubleshooting, code generation, data analysis, and circuit design, as indicated by footnote markers.}
\vspace{0.5ex}
\footnotesize{
\textsuperscript{a} Circuit debugging and troubleshooting \quad
\textsuperscript{b} Code generation or refinement \quad
\textsuperscript{c} Data analysis/library assistance \quad
\textsuperscript{d} Circuit assembly and design
}
\end{table*}

\subsection{Participants}
We recruited fifteen participants in total through our institution's online community post. 
Three were excluded from analysis: who could not understand the task within the allotted time; due to a system malfunction where the AI model failed to surface the required circuit bug; due to a conflict of interest. 
Our analysis, therefore, includes twelve physical computing makers (7 male, 5 female; age=24.92 \(\pm\) 2.84). 
The participants had backgrounds in computer engineering/science ($n=2$), electrical engineering ($n=1$), industrial design ($n=7$), culture technology ($n=1$), and physics ($n=1$); three of them participated in our formative study (P3: formative \sout{study }P1, P9: formative \sout{study }P10, P12: formative \sout{study }P8). 
All participants reported prior education in physical computing, moderate confidence in physical computing (2.9 \(\pm\) 1.5 out of 5-point Likert scale), and similar confidence in software development (3.1 \(\pm\) 1.2 out of 5-point Likert scale). 
All have successfully built more than 2.33 \(\pm\) 1.03 physical computing projects; eleven participants have used a Large Language Model aid for hardware prototyping (Table \ref{tab:usability_participants}).

\subsection{Materials and Method}
Each study session lasted approximately 90 minutes and consisted of five parts: informed consent with demographics survey (10 minutes), system introduction (5 minutes), a circuit building task (55 minutes), post-task evaluations (5 minutes), and a semi-structured interview (15 minutes). 
All participants received compensation equivalent to \$15 USD in local currency.
Two authors were present throughout each session, with one leading the study and providing system tutorials while the other documented participant activities and behaviors for analysis.

\textbf{System Introduction} 
After collecting demographic information, one author introduced participants to \systemName using a prepared software schematic in a live demonstration format. 
Participants were systematically walked through each system feature, including schematic editing, component highlighting, LED blinking functionality, \textit{Ask mode} for conversational guidance, and \textit{Test mode} for circuit validation. 
One author answered any questions about the system, ensuring participants understood all available functionality throughout the construction task. Participants explored the system features freely to familiarize themselves with the interface\sout{ and workflow}.

\begin{figure}[h!]
    \centering
    \includegraphics[width=0.8\linewidth]{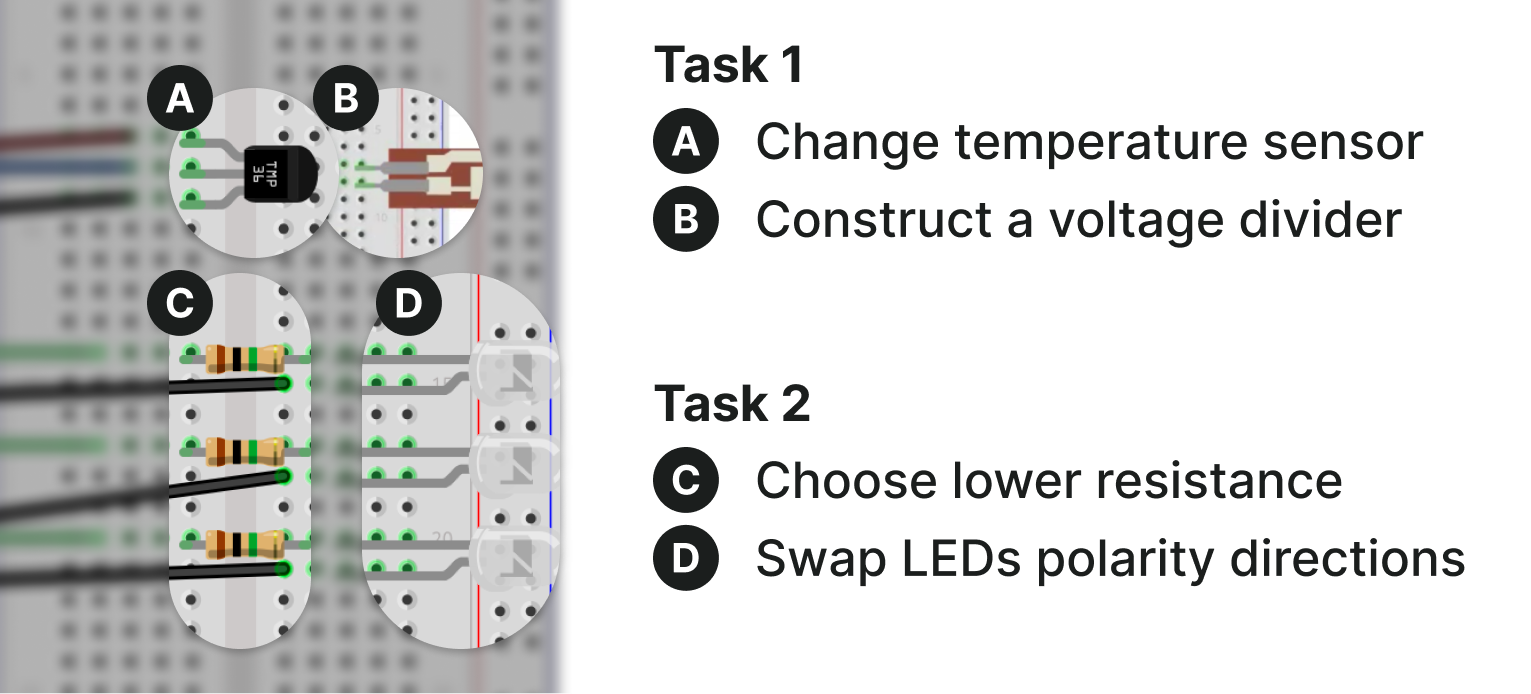}
    \caption{Circuit modification tasks\sout{ for user study}. Task 1 focuses on flex sensor subcircuit\sout{ modifications}: (A) replacing the temperature sensor with a flex sensor, and (B) constructing a voltage divider\sout{ for the flex sensor}. Task 2 addresses LED subcircuit troubleshooting: (C) adjusting resistance values\sout{ for proper operation}, and (D) correcting LED polarity connections.}
    \label{fig:user_study_errors}
    \Description{Breadboard circuit diagram showing four highlighted modification areas: (A) contains a temperature sensor with an annotation 'change temperature sensor', (b) contains a flex sensor with an annotation 'construct a voltage divider', (c) contains three 1MOhm resistors with an annotation 'choose lower resistance', and (d) contains three LEDs with an annotation 'swap LEDs polarity directions'.}
\end{figure}

\textbf{Circuit Construction Task} 
We adapted Project 3: Love-O-Meter from the Arduino Projects Book \cite{fitzgerald2012arduino}, selecting it as a beginner-to-intermediate project to simulate the common maker scenario of discovering an online circuit diagram and adapting it for personal use. This workflow typically requires debugging, component substitution, and circuit modification based on available components and project requirements.
Participants received a partially completed Love-O-Meter circuit within our schematic design interface and were asked to: (1) replace the original temperature sensor with a flex sensor, and (2) identify and fix bugs within the LED circuit (Figure \ref{fig:user_study_errors}). The task was designed to resemble an ecologically valid scenario with participants leveraging \systemName for circuit modification while working with a schematic diagram that partially matches their needs.
Participants could complete these tasks in any order within the allotted timeframe and were not informed about the specific number of bugs to find. We allocated 55 minutes for task completion, providing additional time beyond the original 45-minute recommendation to account for the system learning curve and debugging requirements. To focus the study scope on hardware construction, functional code was provided and accessible throughout the task, and all necessary electronic components were supplied at the beginning of each session.

\textbf{Post-Task Debrief}
Following task completion, participants completed post-task measurements including the System Usability Scale (SUS), NASA Task Load Index (TLX), Trust in Automation (TiA) matrix, and seven custom questionnaires evaluating system capabilities (Appendix \ref{appendix:evaluationq}).
We conducted semi-structured interviews covering task outcomes, prior experience with LLM-assisted hardware prototyping, and system usage observations (Appendix \ref{appendix:userstudy}).

\begin{figure*}[ht!]
    \centering
    \includegraphics[width=0.75\linewidth]{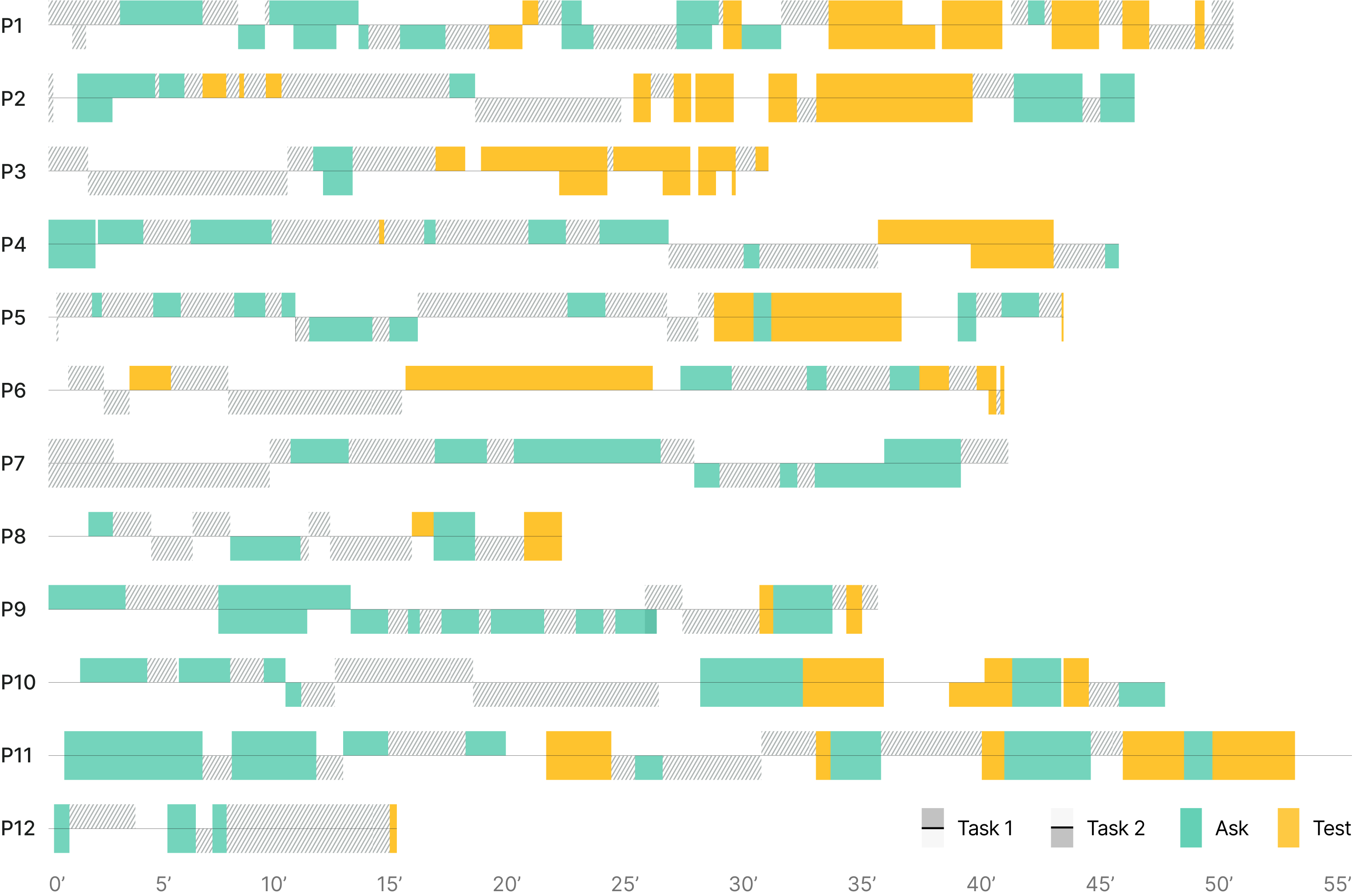}
    \caption{Comprehensive timeline of all participant sessions\sout{ showing system feature usage and task progression}. Each row represents one \sout{participant's }complete session\sout{, with colored blocks indicating different activities}: green (\textit{Ask mode} usage), yellow (\textit{Test mode} usage), gray striped \sout{areas marking }Task 1 \sout{(flex sensor integration) }and Task 2\sout{ (LED circuit debugging) sections}.}
    \label{fig:user-study-timeline}
    \Description{Timeline visualization showing 12 horizontal rows (P1-P12), each representing a participant's session from start to finish. Green blocks indicate Ask mode usage, yellow blocks show Test mode usage, and gray striped areas represent Task 1 (flex sensor integration) and Task 2 (LED circuit debugging) sections.}
\end{figure*}

\subsection{Data Collection and Analysis}
We recorded participant actions through screen capture and external camera footage to document their building process for subsequent analysis. 
We analyzed quantitative data using descriptive statistics and Pearson correlation analysis to examine relationships among experience measures, usability ratings, trust levels, and task performance.
Scale reliability was assessed using Cronbach's alpha, with significance testing at p < 0.05 for correlation analysis.
Interviews were transcribed, translated to English, and analyzed using a general inductive approach with coding consistency verification between two authors \cite{GeneralInductive}. 
When quotes received multiple codes, authors discussed until reaching consensus, prioritizing codes based on user intent and system design relevance.
Following coding completion, we grouped codes into higher-level themes for presentation alongside quantitative results in the subsequent section.

\section{Results and Findings}

Nine out of twelve participants successfully completed working circuits ($37^{\prime}34^{\prime\prime} \pm 12^{\prime}54^{\prime\prime}$, range $14^{\prime}42^{\prime\prime}$-$50^{\prime}00^{\prime\prime}$), with a temporal distribution of system feature usage (\textit{Ask mode}, \textit{Test mode}) and task progression across all participants (Figure \ref{fig:user-study-timeline}).
Three participants failed: P2 and P4 could not design the replacement voltage divider, and P10 used incorrect resistor values due to color band misidentification. 
The system chat response time averaged 8.91 \(\pm\) 5.13 seconds, with maximum response time being 36 seconds and minimum being 2 seconds.
The overall usability of \systemName was evaluated with an average SUS score of 70.4 \(\pm\) 15.2 out of 100, indicating above-average usability \cite{SUS_interpretation}. 
The reported cognitive load from NASA TLX averaged 42.8 \(\pm\) 10 out of 100, representing a moderate workload with balanced mental and physical demands.
TiA scores averaged 3.74 \(\pm\) 1.04 out of 5, indicating moderate to high trust levels. 
All scales demonstrated acceptable to excellent reliability: SUS ($\alpha = .843$), TiA ($\alpha = .82$), and system evaluation questionnaire ($\alpha = .707$).

\begin{figure*}[h!]
    \centering
    \includegraphics[width=1\linewidth]{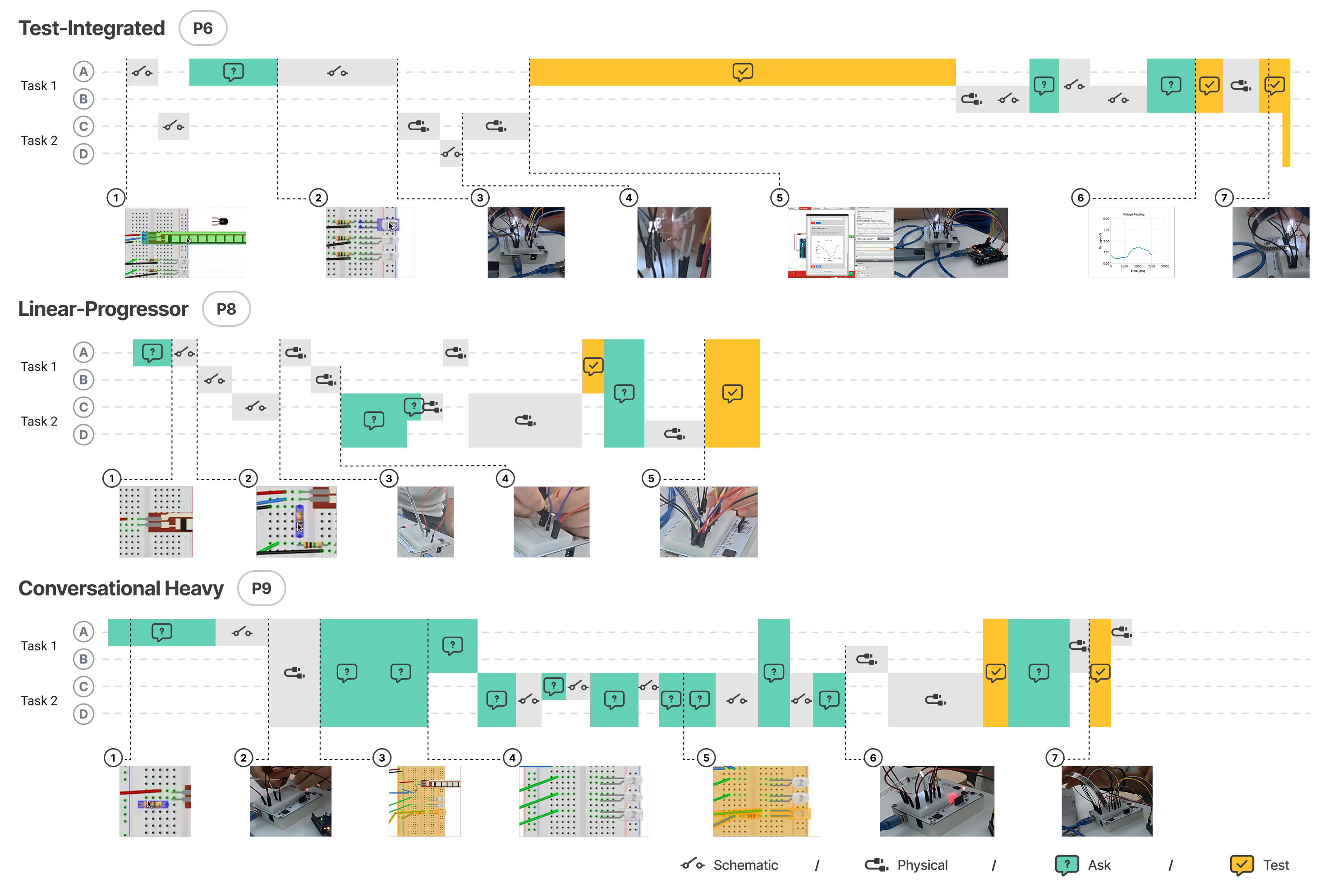}
    \caption{Timeline \sout{visualization }showing distinct workflow patterns of three representative participants. Each row represents one participant's session, with \sout{overlaid} icons \jw{for}\sout{representing} different activities (schematic editing, physical construction, testing, and asking the system).}
    \label{fig:task-screenshot-timeline}
    \Description{The figure shows three horizontal timelines for participants P6, P8, and P9, each demonstrating different workflow archetypes. P6 (test-integrated) shows frequent alternation between construction and testing activities (indicated by checkmark icons) across different colored task blocks, demonstrating iterative validation throughout the session. P8 (linear-progressive) displays sustained activity blocks with minimal testing icons, progressing systematically through the colored task sequence. P9 (conversational-heavy) shows numerous question mark icons distributed throughout various task blocks, indicating heavy reliance on the Ask mode guidance while working across different debugging challenges. Screenshots below each timeline show the corresponding physical circuit states at different stages, illustrating how each participant's circuit evolved during their respective workflows.}
\end{figure*}

\subsection{Diverse Circuit Construction Approaches}
\begin{figure*}[h!]
    \centering
    \includegraphics[width=0.8\linewidth]{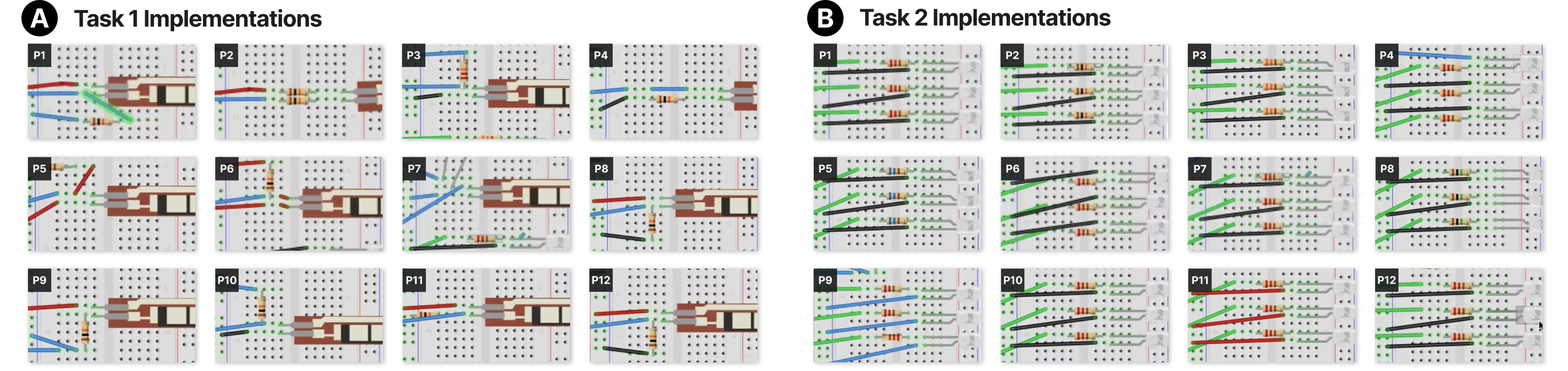}
    \caption{Breadboard circuit implementations by all study participants (N=12), demonstrating individual variation in component placement, wiring strategies, and circuit organization approaches for identical functional requirements. Panel A shows Task 1 implementations, Panel B shows Task 2 implementations.}
    \label{fig:Variety_of_connections_T1}
    \Description{Two panels showing breadboard circuit grids. Panel A displays 12 breadboard circuits labeled P1-P12 showing varied approaches to Task 1, with different component placements, wire routing patterns, and spatial organizations despite identical circuit functionality. Panel B shows the same 12 participants' approaches to Task 2, again demonstrating diverse implementation strategies including different wire colors (red, blue, green, yellow), component orientations, and circuit layouts across the breadboard grid. Each participant's solution is functionally equivalent but visually distinct in organization and construction approach.}
\end{figure*}

Participants demonstrated highly individualized circuit construction approaches with substantial variation in entry points, task progression (Figure \ref{fig:task-screenshot-timeline}), and circuit design (Figure \ref{fig:Variety_of_connections_T1}). 
Behavioral analysis revealed three workflow archetypes: Linear-progressors (P1, P5, P8, P10-P12) maintained longer activity blocks with moderate guidance-seeking; Test-integrated participants (P2, P3, P6) frequently alternated between construction and validation; Conversation heavy participants (P4, P7, P9) relied extensively on \textit{Ask mode} guidance with minimal testing.

\textbf{Task-Switching Frequency} Participants exhibited significant variation in task-switching complexity, with an average of 9.3 \(\pm\) 5.3 activity transitions per session (range 2-17). 
Nine of twelve participants began their session using the Ask feature to gather information about Task 1 or the overall study context, while the remaining three immediately started by editing the schematic, indicating different cognitive entry points into the problem space (Figure \ref{fig:user-study-timeline}). The system's support for idiosyncratic workflows is clearly demonstrated through the distinct color patterns and directional progressions in the timeline data, ultimately resulting in diverse yet functionally equivalent circuit implementations (Figure \ref{fig:Variety_of_connections_T1}). 
The most frequent initial question concerned Task 1 (P3-P10, P12), with a subset of these questions (P7-P8, P10, P12) specifically asking about replacing the temperature sensor with it.

\textbf{Progression Patterns} 
Experienced participants (4+ years: P1, P8, P11) demonstrated non-sequential task progression, moving freely between circuit sections based on debugging insights, with higher task-switching rates (M=0.26 switches/min) compared to novices with under two years of experience (P2, P4, P7; M=0.17 switches/min).
Figure \ref{fig:task-screenshot-timeline} illustrates these patterns across three representative participants. P6 exhibits a test-integrated approach, frequently alternating between construction and testing validation. P8 demonstrates a linear-progressive workflow, systematically progressing through tasks with sustained activity blocks and minimal testing. P9 represents a conversational-heavy archetype, with extensive \textit{Ask mode} usage distributed throughout the timeline.

\textbf{Multilingual Usage} 
Participants demonstrated personalized interaction preferences with both software and hardware components.
Some participants (P10, P12) interacted in their native language without instruction, demonstrating the system's natural language processing capabilities. 
P7 explicitly requested broader language support: \textit{"There's nothing else besides English (that I want to change with the system); I wanted language support other than English."} 

\textbf{Schematic-Hardware Visual Matching}
Participants showed strong preferences for maintaining visual consistency between schematic representations and physical implementations.
P9 appreciated exact connection matching: \textit{"This was very convenient because if I clicked [the component in the schematic software], it blinked. So I could check exactly."}
Some participants even implemented color-coding strategies to maintain visual consistency. When asked about changing wire colors in the software, P11 explained: \textit{"Yeah, so I think it's also related to my working habit. This time I was only able to use the jumper cables from the box, but originally whenever I'm working, I tried to use color coding so that I can select and identify the wire that I'm working on."}

\subsection{Hardware-Contextualized Guidance}
Hardware-software integration proved valuable for reducing the communication overhead typically required with AI assistants on physical computing tasks.
All participants utilized contextual guidance features, sending 166 messages, including 77 with highlighted components for enhanced context.

\textbf{Contextual Component Reference}
All participants used pronouns as references when communicating about circuit components, with 70 pronoun instances across 77 highlighted component interactions (Table \ref{tab:pronouns}).
Most frequent references were \textit{"this"}, \textit{"it"}, and \textit{"here"} with their equivalents, demonstrating reliance on contextual awareness rather than precise component identifier.
Communication logs revealed 8 instances of native language contextual references occurring naturally without multilingual instruction.
In 100\% of component-related communications with component selection, participants used deictic references instead of specific names, indicating significant communication overhead reduction compared to traditional text-based AI assistants.
Examples include P12 asking \textit{"I am not sure where to add this"} while highlighting a $220\Omega$ resistor, and P11 inquiring \textit{"Is it properly connected?"} when referencing a highlighted resistor.

\newcolumntype{M}[1]{>{\centering\arraybackslash}p{#1}}
\newcolumntype{Z}{>{\RaggedRight\arraybackslash\hyphenpenalty=10000\exhyphenpenalty=10000}X}

\begin{table}[h!]
\small
\centering
\caption{Contextual pronoun usage in component-highlighted conversations}
\label{tab:pronouns}

\setlength{\tabcolsep}{4pt}
\renewcommand{\arraystretch}{1.05}

\begin{tabularx}{\linewidth}{l M{2em} M{4.4em} Z}
\toprule
\textbf{Pronoun Type} & \textbf{Uses} & \textbf{Percentage} & \textbf{Examples} \\
\midrule
\enquote{this} \& equivalents & 29 & 41.4\% & \enquote{I am not sure where to add this.} \\
\enquote{it} \& equivalents   & 19 & 27.1\% & \enquote{Is it properly connected?} \\
\enquote{here} \& equivalents & 7  & 10.0\% & \enquote{Can I put the component here?} \\
\enquote{these}                & 5  & 7.1\%  & \enquote{How do I connect these?} \\
\enquote{they/them}            & 4  & 5.7\%  & \enquote{Where should they go?} \\
\enquote{that}                 & 3  & 4.3\%  & \enquote{What does that mean?} \\
\enquote{there}                & 2  & 2.9\%  & \enquote{Put the wire there.} \\
\enquote{now} equivalent       & 1  & 1.4\%  & \enquote{What about now?} \\
Multilingual       & 8  & 11.4\% & Native-language contextual references. \\
\bottomrule
\end{tabularx}
\Description{Frequency analysis table of contextual pronouns used by participants when referencing highlighted circuit components, showing 8 pronoun categories with total usage counts, percentages, and example phrases. "This" and equivalents dominate usage at 41.4\% (29 uses), followed by "it" and equivalents at 27.1\% (19 uses). Together, these two categories account for 68.5\% of all contextual references. Spatial reference pronouns include "here" (10.0\%), "these" (7.1\%), "they/them" (5.7\%), "that" (4.3\%), "there" (2.9\%), and temporal "now" equivalents (1.4\%). Additionally, 11.4\% of references used multilingual equivalents in participants' native languages. Examples demonstrate natural conversational patterns such as "I am not sure where to add this" and "Is it properly connected?" indicating participants' reliance on visual highlighting context rather than technical component names.} 
\end{table}

\textbf{Eliminating Manual Circuit Explanation}
The system's circuit state awareness eliminated extensive verbal descriptions typically required for general-purpose AI assistants. 
P2 highlighted this advantage: \textit{"It is very uncomfortable using ChatGPT [to manually say about the context all the time]. I think that is the advantage of the system that I can just add context by clicking the button, and I don't need to explain \sout{the }details of pins or preferences."} 
P7 emphasized: \textit{"Using an assistant like this means I don't have to explain to GPT how the circuit is structured, and that's the difference."}
Contextual awareness from schematic updates and component highlighting enabled automatic error detection without explicit requests.
P8 appreciated, \textit{"\sout{For example, }I don't remember the color of each resistor, which color means which value, but \sout{the }system automatically found that I used the wrong resistor... that automatic thing was quite convenient for me."}

\textbf{Direct Hardware Guidance Through LED}
The component blinking on the breadboard demonstrated high adoption rates and significant correlations with system trust measures. 
All participants except P12 utilized the Blink feature, generating 297 total highlighting interactions across sessions.
The system supported three distinct highlighting modalities: self-initiated position verification ($n=84$), AI-guided blinking ($n=11$), and test-integrated highlighting ($n=32$).
Position-based highlighting had the highest adoption, with 11 of 12 participants using it to verify component locations. 
Usage patterns varied significantly: P2 engaged the highest (21 position clicks, 34 total interactions) while P12 showed no usage.
P12 explained: \textit{"Checking whether resistors or LEDs are properly connected in a straight line on a specific row seems more important."}, preferring manual connection checking.
P2 contrasted: \textit{"I used to get lost in wiring the board, but the visual assistant is very comfortable."}
System-guided highlighting was used by only 3 participants (P1, P2, P5), while test-integrated highlighting was adopted by 7 participants. 
Strong correlations emerged between highlighting usage and system trust: AI-guided blinking users reported lower system malfunction concerns ($r = -0.71$, $p = .009$), while frequent position verification users demonstrated higher confidence in system capabilities ($r = 0.62$, $p = .030$).

\textbf{Validation of AI Recommendations}
LED highlighting spatial guidance received widespread appreciation, with participants finding visual cues more intuitive than text-based instructions alone, supporting both autonomous exploration and system-guided assistance.
Participants appreciated contextual awareness but remained appropriately critical of AI suggestions. 
P3 exemplified this balance: \textit{"I did not trust the chatbot with the actual value of this pressure sensor... And I was kind of right. My gut feeling was right because in the end, we saw that I had to change the resistor a couple of times."} 
Participants who better understood system reasoning found schematic-board action connections clearer ($r = 0.78$, $p = .003$). 
Those appreciating the system's complex task handling were less concerned about potential errors ($r = -0.77$, $p = .003$).
P3 described: \textit{"The LED highlighting was pretty nice... the most helpful thing... if I just had a schematic, it would be an initially much more overwhelming task to start."}

\subsection{Ad Hoc Test Generation for Circuit}

The testing capabilities proved highly popular, with 7 out of 12 participants actively using \textit{Test mode}, demonstrating the system's ability to automatically generate appropriate tests based on schematic\sout{ awareness}.

\textbf{Simplified Testing Setup}
Participants valued testing without code modifications or complex hardware reconfiguration (P2-P10). 
P6 explained: \textit{"I also like the test parts, so I don't need to re-upload test codes."} 
P7 emphasized workflow benefits: \textit{"When testing a set, if you're working alone and just building something, you'd normally have to remove it, make a test setup, try it again, then upload it. But with this, you just plug in one wire, and that's it."} 
P3 added: \textit{"[I want to have more test types] to be platform specific ... for sweeping the voltage for the analog input pin."}
Contextual awareness of circuit topology enabled automatic test procedure generation. The system eliminated overhead from creating additional testing setups with multimeters and writing dedicated testing code by understanding component testing needs and automatically suggesting relevant measurement approaches.

\begin{figure}[h!]
    \centering
    \includegraphics[width=1\linewidth]{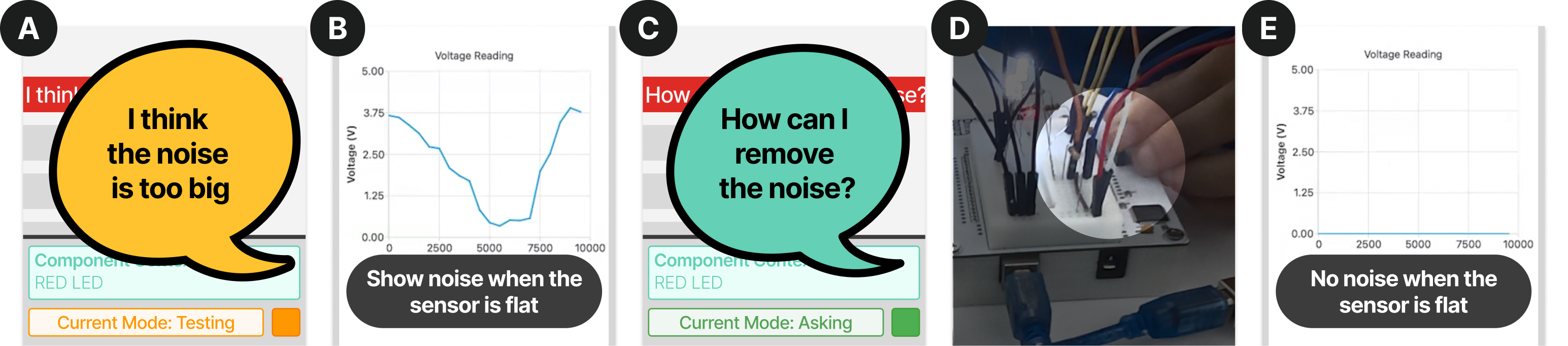}
    \caption{P6's \sout{systematic }circuit debugging workflow\sout{ using \systemName's testing capabilities. The process} demonstrates: (A) initial problem identification reporting excessive sensor noise, (B) conducting system-suggested diagnostic tests to measure flex sensor noise levels, (C) requesting solution guidance\sout{ from the AI agent}, (D) implementing the recommended circuit modification\sout{ by adding a resistor to the flex sensor subcircuit}, and (E) verification tests to confirm noise reduction.}
    \label{fig:testing-noise-problem}
    \Description{Five-panel workflow sequence showing P6's debugging process: Panel A displays chat interface with noise complaint, Panel B shows voltage measurement graph with high noise readings, Panel C shows AI agent conversation requesting solution, Panel D shows physical breadboard with resistor being added to circuit, and Panel E displays updated measurement graph showing reduced noise levels.}
\end{figure}

\textbf{Component-Specific and Whole-Circuit Testing}
The system's contextual awareness further streamlined testing by understanding circuit context without manual explanation. 
P6 appreciated not having to explain \sout{their }setup textually nor visually with pictures: \textit{"I usually use ChatGPT and I ask it to make some skeleton calls when I have some error on the code, I asked AI to fix it. And sometimes when the circuits, I think when the error comes from the circuit, I take pictures and let AI figure out what."} 
The ability to test individual components proved invaluable for problem isolation, as demonstrated in Figure \ref{fig:testing-noise-problem}, where P6 systematically identified and resolved noise issues with the flex sensor.
P6 expressed a desire for even smarter automation: \textit{"I want to say I want to test this sensor and it can automatically switch to \textit{Test mode} and test for me."}
Strong correlations emerged between testing effectiveness and overall system trust, with participants who found tests helpful for fault isolation showing higher confidence in system capabilities ($r = 0.67$, $p = .017$).

\textbf{Test Results Analysis}
Participants who could distinguish reliable from inconclusive test results showed stronger correlations with system-suggested measurement accuracy ($r=0.79$, $p = .002$), indicating that effective result interpretation was crucial for leveraging the testing capabilities.
However, during the testing experience, several participants struggled with interface confusion and inconsistent results. 
P2 noted: \textit{"I am confused about the ability of \textit{Ask mode} and testing mode because I didn't spot the major difference."} 
P6 experienced frustration when testing failed: \textit{"I tried adding the pulldown resistor, but I failed actually because ChatGPT recommended me the wrong resistor value. And I tried to use the serial monitor to see the noise for the flex sensor, but it was not a very good experience."} 
This confusion led some participants (P8, P11) to bypass \textit{Test mode} entirely, preferring the serial monitor instead.
Despite these interface challenges, the underlying functionality successfully supported debugging workflows when participants understood how to interpret the results. 
Additional correlations demonstrated that participants who found measurement suggestions accurate were more confident in system capabilities ($r = 0.64$, $p = .024$), and those who understood test outcomes showed higher system reliability ratings ($r = 0.66$, $p = .020$).

\subsection{System Reliability and Usage Context Feedback}

\textbf{System Trust and Acceptance}
Strong correlations emerged between system acceptance and trust mechanisms, revealing specific relationships relevant to system design.
Participants' intention for frequent use was strongly associated with system reliability ($r = 0.88$, $p < .001$), while those who found system functions well-integrated showed reduced unpredictability concerns ($r = -0.83$, $p < .001$). 
Prior AI experience showed no relationship with task performance ($r = 0.00$, $p = .994$), while physical computing experience did not significantly predict system evaluation ratings ($r = 0.25$, $p = .440$), indicating \systemName successfully accommodated users regardless of technical background. 

\textbf{Educational and Professional Applications}
Interview analysis revealed distinct usage patterns across educational and professional contexts.
Most participants identified educational potential with conditions. 
P3 explained: \textit{"I would really see this as at a pre-university electronics courses, such as in science high school or Robotics classes... especially teenagers would love this idea."} 
However, P9 countered: \textit{"I would not recommend using this project if you're a kid and you want to learn hardware, because... they tell us everything, and you just copy what it is. So at the end of the day, you end up learning nothing."} 
This highlights the importance of balancing assistance with learning opportunities.
Beyond education, participants identified value for non-engineers in creative projects. 
P3 and P11 noted applications for interaction artists, with P11 mentioning: \textit{"when they don't know about the detailed technical aspect. They will find it really helpful."} 
The system's value extended to personal project development. 
P6 exemplified, explaining how the system solved her flex sensor project: 
\textit{"I am using this flex sensor for my personal project now, and this sensor was a little tricky because I cannot see what's happening and why the noise is huge. And I tried the same method, adding the pulldown resistor, but I failed... ChatGPT recommended me the wrong resistor value."} 
Through guided testing, she learned that the flex sensor required a proper voltage divider configuration. 
P6 appreciated reduced trial-and-error: \textit{"was really helpful for educational things because before I went through many trials and errors, I would make so many mistakes, like name changes, wrong resistors, or broken circuits. So it's beneficial because it firstly makes the circuits in the software and then goes to here [breadboard], and I really like the blinking for where you should put the wire, because sometimes I think that I did it correctly, but actually I did it wrong."} 
This demonstrates how \systemName bridged formal learning in practical applications.

\textbf{Circuit Complexity Considerations}
Participants had mixed perspectives on complex circuit usage. 
Multiple participants indicated they would not use the system for highly complex circuits.  
P9 emphasized utility for smaller components: \textit{"if it's a small part of the task of the bigger picture, I think it's fine."} also P10 explained, \textit{"It'd not be useful for people who've already designed so much [because then they need to add all of their circuit design in software], would be better just to casually make from the beginning."} 
However, some participants suggested usefulness for complex circuits, noting potential applications in custom PCB design.
P8 mentioned: \textit{"I can imagine also having a custom PCB board to use with this system, as I can just upload that PCB file to use the system."}
This suggests diverse application opportunities while highlighting the need for improved interface clarity and assistance-learning balance.

\section{Discussion}

The formative study highlighted three key challenges among makers for physical computing prototyping, leading to our three design goals: DG1) personalized guidance adapting to individual circuit contexts; DG2) hardware-contextualized visual guidance bridging schematics with physical construction; and DG3) context-aware testing generating validation procedures for specific configurations.
All participants adopted these core capabilities, revealing distinct workflow archetypes through personalized approaches, widespread use of pronoun-based contextual communication over technical terminology, and effective hardware-aware debugging. Participants valued \systemName's usefulness while noting mode confusion and concerns about balancing assistance with learning opportunities in educational contexts.

\subsection{Physical Circuits as Spatial Language}

Our approach presents hardware-software interaction as a communication problem, building on \citet{Miyake_1986}'s work explaining that conceptual viewpoints are reflected in \textit{spatial language} during collaborative understanding of complex physical devices.
The distinctive contribution lies in combining real-time schematic parsing with conversational interaction to support makers' idiosyncratic workflows \cite{CrossedWires} from arbitrary starting points.
This addresses the formative study's first challenge of diverse construction strategies, as evidenced by the distinct workflow archetypes (test-integrated, linear-progressive, conversational-heavy) that emerged during evaluation.
Addressing the second challenge of schematic-physical mismatch, \systemName maintains awareness of exact circuit topology rather than requiring users to manually bridge between representations while enabling natural spatial communication patterns. 
This dual capability of understanding both the precise technical state and the conversational context enables support for individual preferences and non-linear exploration that characterizes authentic maker practices \cite{McLaughlin_2025}. 
The key insight is that when systems understand circuit context, users intuitively abandon technical identifiers in favor of deictic references, mirroring a natural communication strategy of humans in physical spaces.

Previous hardware guidance approaches span instruction delivery systems \cite{BlinkBoard, SchemaBoard, VisibleBreadboard}, conversational interfaces for step-by-step guidance \cite{HeyTeddy, FritzBot}, structured tutorial frameworks \cite{Ritschel_2023, ElectroTutor}, and example-based exploration tools \cite{Muse}. Our communication paradigm builds on foundational work in constructive interaction \cite{Miyake_1986} and conversation with materials \cite{Torres_2019}, yet differs \sout{fundamentally }by combining real-time schematic parsing with contextual communication. 
While existing tools assume predetermined workflows, \systemName combines technical analysis and natural dialogue from arbitrary circuit origins.
This distinction allows complementary use: providing contextual guidance wherever users begin without disrupting exploration patterns, alongside structured instruction systems.

Our work suggests a fundamental design principle: automated context understanding enables personalized communication. 
Systems that parse domain-specific state can support natural human communication patterns rather than forcing conformity to predetermined workflows. 
For HCI, this implies combining deep technical awareness with adaptive interaction paradigms that accommodate diverse user preferences, from multimodal interfaces supporting different sensory modalities \cite{IncluSim} to tutorial mediums that affect physical skill transfer \cite{Endow_2021}.
By supporting inquiries about arbitrary component models rather than requiring predetermined sets, \systemName enables makers to utilize existing materials, potentially supporting sustainable making practices through component reuse and reduced e-waste in prototyping activities \cite{ProtoPCB, Yan_2025}.
These findings extend beyond electronics to broader physical manipulation concerns \cite{MeasurementPatterns}, contributing to physical computing's foundational goals \cite{physicalcomputing} through intelligent mediation rather than rigid instruction. 

\subsection{Debugging Circuits as an Inquiry}
Addressing the third challenge of manual and iterative debugging, we reframed circuit debugging from a diagnostic skill to a collaborative inquiry process, building upon Interrogative Debugging \cite{Whyline}, where programmers ask "why did" and "why didn't" questions about program failures.
Instead of requiring users to hypothesize error sources and manually construct test procedures \cite{DeLiema_2024}, our approach treats the circuit itself as an active participant in the debugging conversation. This shifts the cognitive burden from error localization to result interpretation, democratizing debugging capabilities while preserving learning opportunities \cite{Hennessy_2023}.

Physical computing debugging approaches typically require significant technical expertise \cite{Bifrost, Heimdall}, explicit hypothesis formulation \cite{Hypothesizer}, predefined testing frameworks \cite{ElectroTutor}, or manual interpretation of visualization data \cite{CurrentViz, CircuitSense}. Software-focused advances \cite{Inline, WiFrost} and real-time manipulation tools \cite{Scanalog} with structured logging approaches \cite{LogIt} operate independently from circuit construction contexts. Our collaborative inquiry approach, inspired by foundational "why" questioning work \cite{Ko_2008, Whyline}, transforms debugging from diagnostic skill to guided exploration. Educational research \cite{Hennessy_2024} emphasizes diverse debugging pathways, aligning with our democratization goals. \systemName complements existing tools by providing circuit-aware test generation that adapts to any hardware configuration, reducing expertise barriers while preserving learning value and maintaining appropriate trust in automated systems \cite{TrustinAutomation}.

Our findings reveal that effective debugging tools should function as mediators of inquiry rather than diagnostic instruments, suggesting a design philosophy where tools scaffold exploration rather than prescribe solutions. This points toward debugging interfaces that emphasize collaborative sense-making between human intuition and computational analysis, with broader implications for any domain requiring iterative refinement of complex systems. 
However, three critical obstacles emerge for LLM-based assistance in maker contexts. 
First, over-reliance risks may undermine the productive struggle essential for developing debugging expertise \cite{Hennessy_2023, DebugbyDesign}, a concern increasingly recognized in educational AI systems \cite{chu-etal-2025-llm}. 
Next, while LLMs help bridge vocabulary gaps, the system may respond with technical terminology (e.g., cathode) that novices may struggle to translate without visual or contextual support. 
Lastly, this challenge could be compounded by cultural and dialectal biases in language models \cite{Wasi_2025} that may exclude non-English speaking makers. 
Future tools could balance automation with agency through adaptive support that fades as competence develops \cite{Ritschel_2023}, while addressing language barriers to ensure users remain active participants rather than passive observers in troubleshooting processes.

\subsection{Limitations and Future Works}

We acknowledge \sout{that }\systemName's current implementation presents software and hardware limitations that offer opportunities for enhancement.

\textbf{Software Design:} While the \textit{Ask/Test mode} separation functions as a fail-safe, this distinction created confusion among participants, suggesting mode selection based on user query and conversational context.
Enhanced AI prompt pipelines could infer user intent by simultaneously processing natural language input, schematic state, breadboard connections~\citep[e.g.,][]{CircuitSense}, and code environments.
Added to current pin highlighting for voltage outputs, future systems could introduce explicit feedback mechanisms: mode clarification through confirmation dialogs when users switch modes (e.g., "Proceed for testing?") to help users understand system state, and safety warnings before voltage outputs (e.g., "Outputting 5V at A0. Proceed?") to prevent component damage and encourage safety practices \cite{Lahaye_2023}.

\textbf{Hardware State:} Our system cannot physically sense component presence, identity, or connection topology unlike CircuitSense \cite{CircuitSense} or CurrentViz \cite{CurrentViz}. 
Users still need to verify the synchronization manually to ensure the schematics match those on the breadboard, compounded by LED visibility issues, as participants reported eye strain and difficulty distinguishing specific breadboard rows, particularly under varying lighting conditions.
Future works could explore automated circuit sensing or alternative indication methods such as directional lighting, improved LED diffusion, or augmented reality overlays \cite{SpatIO, VirtualComponent} that could both guide component placement and assist users in verifying circuit state.
The pins can output basic electronic signals (e.g., analog voltage, PWM), and read analog voltage. Subsequent iterations can integrate more advanced input/output protocols, like UART, I2C, or SPI, to assist testing digital components.
The 25-row breadboard constrains the complexity of circuits. Future hardware iterations could house a larger breadboard or enable multi-board configurations via communication ports. Software could also support virtual component placement beyond the physical board \cite{VirtualComponent}.

\textbf{Evaluation Methodology:} Our usability study introduces three constraints that limit generalizability and comparative assessment with other systems.
Firstly, the study involved simple circuits with moderate component density, limiting transferability to more complex configurations. Future studies should investigate more complex circuit configurations in longitudinal settings to examine skill transfer and retention over extended periods.
Second, the absence of baseline comparisons prevents direct quantification of \systemName's advantages over traditional or existing tools like \cite{HeyTeddy, FritzBot}.
Additionally, the formative study restricted participants from using LLMs beyond Google's default AI summaries to establish baseline challenges, while \systemName's core support derives from LLMs. Further studies should incorporate controlled comparisons with conventional approaches, established tutorial systems, general-purpose AI assistants, and separate tool combinations (e.g., LLMs with standalone schematic design software) to validate the integration value. 
Third, participant recruitment focused primarily on university students with moderate physical computing experience, limiting insights about effectiveness across diverse skill levels and cultural contexts. 
Expanding evaluation to include professional makers, educators, and participants from varied educational backgrounds would strengthen claims about universal applicability. 

\section{Conclusion}

We presented \systemName, an integrated development environment that bridges circuit design software and physical construction, through hardware-contextualized guidance and in-situ testing capabilities. 
It allows makers to interact with an AI that maintains real-time awareness of both visual circuit schematics and physical configurations, supported by LED-based spatial guidance on an augmented breadboard and automatically generated context-aware tests. 
The system accommodates personalized building workflows by enabling natural contextual references to circuit components and eliminating the need for separate testing setups. 
The system was designed based on formative studies with makers and was evaluated with 12 participants across circuit construction tasks. 
Results show that participants successfully adopted all main features, with three distinct workflow archetypes emerging while maintaining high completion rates and universal accessibility across technical backgrounds. 
Future work will explore automatic mode change through enhanced prompt pipelines, hardware improvements such as circuit sensing and augmented reality overlays, and longitudinal evaluations with diverse maker populations.

\begin{acks}
This work was supported by the IITP(Institute of Information \& Communications Technology Planning \& Evaluation)-ITRC(Inform-
ation Technology Research Center) grant funded by the Korea government(Ministry of Science and ICT)(\mbox{IITP-2026-RS-2024-00436398}). 
\end{acks}

\bibliographystyle{ACM-Reference-Format}
\bibliography{references}

@ARTICLE{physicalCSE,
  author={Hodges, Steve and Sentance, Sue and Finney, Joe and Ball, Thomas},
  journal={Computer}, 
  title={Physical Computing: A Key Element of Modern Computer Science Education}, 
  year={2020},
  volume={53},
  number={4},
  pages={20-30},
  doi={10.1109/MC.2019.2935058},
  ISSN={1558-0814},
  month={April},}

@inproceedings{Wasi_2025,
author = {Wasi, Azmine Toushik and Islam, Raima and Islam, Mst Rafia and Sadeque, Farig and Rafi, Taki Hasan and Chae, Dong-Kyu},
title = {Dialectal Bias in Bengali: An Evaluation of Multilingual Large Language Models Across Cultural Variations},
year = {2025},
isbn = {9798400713316},
publisher = {Association for Computing Machinery},
address = {New York, NY, USA},
url = {https://doi.org/10.1145/3701716.3715468},
doi = {10.1145/3701716.3715468},
booktitle = {Companion Proceedings of the ACM on Web Conference 2025},
pages = {1380–1384},
numpages = {5},
location = {Sydney NSW, Australia},
series = {WWW '25}
}

@inproceedings{chu-etal-2025-llm,
    title = "{LLM} Agents for Education: Advances and Applications",
    author = "Chu, Zhendong  and
      Wang, Shen  and
      Xie, Jian  and
      Zhu, Tinghui  and
      Yan, Yibo  and
      Ye, Jingheng  and
      Zhong, Aoxiao  and
      Hu, Xuming  and
      Liang, Jing  and
      Yu, Philip S.  and
      Wen, Qingsong",
    editor = "Christodoulopoulos, Christos  and
      Chakraborty, Tanmoy  and
      Rose, Carolyn  and
      Peng, Violet",
    booktitle = "Findings of the Association for Computational Linguistics: EMNLP 2025",
    month = nov,
    year = "2025",
    address = "Suzhou, China",
    publisher = "Association for Computational Linguistics",
    url = "https://aclanthology.org/2025.findings-emnlp.743/",
    doi = "10.18653/v1/2025.findings-emnlp.743",
    pages = "13782--13810",
    ISBN = "979-8-89176-335-7"
}

@inproceedings{IncluSim,
author = {Mouallem, Aya and Mendez Pons, Mirelys and Malik, Ali and Rogando, Trini and Kim, Gene S-H and Kulkarni, Trisha and Chong, Charlene and Fan, Danyang and Patel, Shloke Nirav and Shluzas, Lauren Aquino and Chen, Helen L. and Sheppard, Sheri D.},
title = {IncluSim: An Accessible Educational Electronic Circuit Simulator for Blind and Low-Vision Learners},
year = {2025},
isbn = {9798400713941},
publisher = {Association for Computing Machinery},
address = {New York, NY, USA},
url = {https://doi.org/10.1145/3706598.3713437},
doi = {10.1145/3706598.3713437},
booktitle = {Proceedings of the 2025 CHI Conference on Human Factors in Computing Systems},
articleno = {338},
numpages = {18},
location = {
},
series = {CHI '25}
}

@inproceedings{Endow_2021,
author = {Endow, Shreyosi and Torres, Cesar},
title = {“I’m Better Off on my Own”: Understanding How a Tutorial’s Medium Affects Physical Skill Development},
year = {2021},
isbn = {9781450384766},
publisher = {Association for Computing Machinery},
address = {New York, NY, USA},
url = {https://doi.org/10.1145/3461778.3462066},
doi = {10.1145/3461778.3462066},
booktitle = {Proceedings of the 2021 ACM Designing Interactive Systems Conference},
pages = {1313–1323},
numpages = {11},
location = {Virtual Event, USA},
series = {DIS '21}
}

@inproceedings{Lahaye_2023,
author = {Lahaye, Marcel and Reinartz, Vivian Isabel and Sahabi, Sarah and Borchers, Jan},
title = {Towards Authoring Tools For DIY Tutorials: From Tutorial User Strategies to Guidelines (Free Template Included!)},
year = {2023},
isbn = {9798400707711},
publisher = {Association for Computing Machinery},
address = {New York, NY, USA},
url = {https://doi.org/10.1145/3603555.3608530},
doi = {10.1145/3603555.3608530},
booktitle = {Proceedings of Mensch Und Computer 2023},
pages = {380–386},
numpages = {7},
location = {Rapperswil, Switzerland},
series = {MuC '23}
}

@book{schön1983reflective,
  title={The Reflective Practitioner: How Professionals Think in Action},
  author={Sch{\"o}n, Donald A.},
  isbn={9781857423198},
  lccn={82070855},
  series={An Ashgate book},
  url={https://books.google.co.kr/books?id=E85qAAAAMAAJ},
  year={1983},
  publisher={Ashgate}
}

@inproceedings{Yan_2025,
author = {Yan, Zeyu and Dhaygude, Mrunal Sanjay and Peng, Huaishu},
title = {Make Making Sustainable: Exploring Sustainability Practices, Challenges, and Opportunities in Making Activities},
year = {2025},
isbn = {9798400713941},
publisher = {Association for Computing Machinery},
address = {New York, NY, USA},
url = {https://doi.org/10.1145/3706598.3713665},
doi = {10.1145/3706598.3713665},
booktitle = {Proceedings of the 2025 CHI Conference on Human Factors in Computing Systems},
articleno = {886},
numpages = {14},
location = {
},
series = {CHI '25}
}

@inproceedings{ProtoPCB,
author = {Lu, Jasmine and Boddu, Sai Rishitha and Lopes, Pedro},
title = {ProtoPCB: Reclaiming Printed Circuit Board E-waste as Prototyping Material},
year = {2025},
isbn = {9798400713941},
publisher = {Association for Computing Machinery},
address = {New York, NY, USA},
url = {https://doi.org/10.1145/3706598.3714095},
doi = {10.1145/3706598.3714095},
booktitle = {Proceedings of the 2025 CHI Conference on Human Factors in Computing Systems},
articleno = {888},
numpages = {12},
location = {
},
series = {CHI '25}
}

@article{Hennessy_2023,
    author = {Hennessy Elliott, Colin and Gendreau Chakarov, Alexandra and Bush, Jeffrey B. and Nixon, Jessie and Recker, Mimi},
    title = {Toward a debugging pedagogy: helping students learn to get unstuck with physical computing systems},
    journal = {Information and Learning Sciences},
    volume = {124},
    number = {1-2},
    pages = {1-24},
    year = {2023},
    month = {01},
    issn = {2398-5348},
    doi = {10.1108/ILS-03-2022-0051},
    url = {https://doi.org/10.1108/ILS-03-2022-0051},
    eprint = {https://www.emerald.com/ils/article-pdf/124/1-2/1/1147515/ils-03-2022-0051.pdf},
}

@article{Ritschel_2023,
author = {Ritschel, Nico and Sawant, Anand Ashok and Weintrop, David and Holmes, Reid and Bacchelli, Alberto and Garcia, Ronald and K R, Chandrika and Mandal, Avijit and Francis, Patrick and Shepherd, David C.},
title = {Training industrial end-user programmers with interactive tutorials},
journal = {Software: Practice and Experience},
volume = {53},
number = {3},
pages = {729-747},
doi = {https://doi.org/10.1002/spe.3167},
url = {https://onlinelibrary.wiley.com/doi/abs/10.1002/spe.3167},
eprint = {https://onlinelibrary.wiley.com/doi/pdf/10.1002/spe.3167},
year = {2023}
}

@inproceedings{Muse,
author = {Methfessel, Paul and Beckmann, Tom and Rein, Patrick and Ramson, Stefan and Hirschfeld, Robert},
title = {MµSE: Supporting Exploration of Software-Hardware Interactions Through Examples},
year = {2024},
isbn = {9798400703300},
publisher = {Association for Computing Machinery},
address = {New York, NY, USA},
url = {https://doi.org/10.1145/3613904.3642186},
doi = {10.1145/3613904.3642186},
booktitle = {Proceedings of the 2024 CHI Conference on Human Factors in Computing Systems},
articleno = {936},
numpages = {16},
location = {Honolulu, HI, USA},
series = {CHI '24}
}

@book{physicalcomputing,
author = {O'Sullivan, Dan and Igoe, Tom},
title = {Physical computing: sensing and controlling the physical world with computers},
year = {2004},
isbn = {159200346X},
publisher = {Course Technology Press},
address = {Boston, MA, USA}
}

@Inbook{GroundedTheory,
author={Vollstedt, Maike
and Rezat, Sebastian},
editor={Kaiser, Gabriele
and Presmeg, Norma},
title={An Introduction to Grounded Theory with a Special Focus on Axial Coding and the Coding Paradigm},
bookTitle={Compendium for Early Career Researchers in Mathematics Education},
year={2019},
publisher={Springer International Publishing},
pages={81--100},
isbn={978-3-030-15636-7},
doi={10.1007/978-3-030-15636-7_4},
url={https://doi.org/10.1007/978-3-030-15636-7_4}
}

@inproceedings{SpatIO,
author = {Han, Seung Hyeon and Han, Yeeun and Park, Kyeongho and Lee, Sangjun and Lee, Woohun},
title = {SpatIO: Spatial Physical Computing Toolkit Based on Extended Reality},
year = {2025},
isbn = {9798400713941},
publisher = {Association for Computing Machinery},
address = {New York, NY, USA},
url = {https://doi.org/10.1145/3706598.3713747},
doi = {10.1145/3706598.3713747},
booktitle = {Proceedings of the 2025 CHI Conference on Human Factors in Computing Systems},
articleno = {978},
numpages = {22},
location = {
},
series = {CHI '25}
}

@inproceedings{Hypothesizer,
author = {Alaboudi, Abdulaziz and Latoza, Thomas D.},
title = {Hypothesizer: A Hypothesis-Based Debugger to Find and Test Debugging Hypotheses},
year = {2023},
isbn = {9798400701320},
publisher = {Association for Computing Machinery},
address = {New York, NY, USA},
url = {https://doi.org/10.1145/3586183.3606781},
doi = {10.1145/3586183.3606781},
booktitle = {Proceedings of the 36th Annual ACM Symposium on User Interface Software and Technology},
articleno = {73},
numpages = {14},
location = {San Francisco, CA, USA},
series = {UIST '23}
}

@incollection{King_2017,
title = {Chapter 8 - Cross Institutional Peer Coaching: A Case Study},
editor = {Dawn Lowe-Wincentsen},
booktitle = {Beyond Mentoring},
publisher = {Chandos Publishing},
pages = {93-106},
year = {2017},
isbn = {978-0-08-101294-9},
doi = {https://doi.org/10.1016/B978-0-08-101294-9.00008-8},
url = {https://www.sciencedirect.com/science/article/pii/B9780081012949000088},
author = {M. King and D. Winn}
}

@article{Alessandrini_2022,
author = {Alessandrini, Andrea},
year = {2022},
month = {10},
pages = {1-18},
title = {A Study of Students Engaged in Electronic Circuit Wiring in an Undergraduate Course},
volume = {32},
journal = {Journal of Science Education and Technology},
doi = {10.1007/s10956-022-09994-9}
}

@article{Miyake_1986,
title = {Constructive interaction and the iterative process of understanding},
journal = {Cognitive Science},
volume = {10},
number = {2},
pages = {151-177},
year = {1986},
issn = {0364-0213},
doi = {https://doi.org/10.1016/S0364-0213(86)80002-7},
url = {https://www.sciencedirect.com/science/article/pii/S0364021386800027},
author = {Naomi Miyake}
}

@article{DebugbyDesign,
year = {2021}, 
title = {Growing Mindsets: Debugging by Design to Promote Students’ Growth Mindset Practices in Computer Science Class.}, 
url = {https://par.nsf.gov/biblio/10309425}, 
journal = {Proceedings of the 15th International Conference of the Learning Sciences - ICLS 2021}, 
author={Morales-Navarro, Luis and Fields, Deborah A and Kafai, Yasmin B}, 
editor = {de Vries, E. and Hod, Y. and Ahn, J.} 
}

@INPROCEEDINGS{Yusuf_2023,
  author={Bani Yusuf, Imam Nur and Binte Abdul Jamal, Diyanah and Jiang, Lingxiao},
  booktitle={2023 IEEE/ACM 20th International Conference on Mining Software Repositories (MSR)}, 
  title={Automating Arduino Programming: From Hardware Setups to Sample Source Code Generation}, 
  year={2023},
  volume={},
  number={},
  pages={453-464},
  doi={10.1109/MSR59073.2023.00069},
  ISSN={2574-3864},
  month={May},}

@article{DeLiema_2024,
author = {DeLiema, David and Bye, Jeffrey K. and Marupudi, Vijay},
title = {Debugging Pathways: Open-Ended Discrepancy Noticing, Causal Reasoning, and Intervening},
year = {2024},
issue_date = {June 2024},
publisher = {Association for Computing Machinery},
address = {New York, NY, USA},
volume = {24},
number = {2},
url = {https://doi.org/10.1145/3650115},
doi = {10.1145/3650115},
journal = {ACM Trans. Comput. Educ.},
month = may,
articleno = {28},
numpages = {34}
}

@inproceedings{BitBlox,
author = {DesPortes, Kayla and Anupam, Aditya and Pathak, Neeti and DiSalvo, Betsy},
title = {BitBlox: A Redesign of the Breadboard},
year = {2016},
isbn = {9781450343138},
publisher = {Association for Computing Machinery},
address = {New York, NY, USA},
url = {https://doi.org/10.1145/2930674.2930708},
doi = {10.1145/2930674.2930708},
booktitle = {Proceedings of the The 15th International Conference on Interaction Design and Children},
pages = {255–261},
numpages = {7},
location = {Manchester, United Kingdom},
series = {IDC '16}
}

@inproceedings{DesPortes_2019,
author = {DesPortes, Kayla and DiSalvo, Betsy},
title = {Trials and Tribulations of Novices Working with the Arduino},
year = {2019},
isbn = {9781450361859},
publisher = {Association for Computing Machinery},
address = {New York, NY, USA},
url = {https://doi.org/10.1145/3291279.3339427},
doi = {10.1145/3291279.3339427},
booktitle = {Proceedings of the 2019 ACM Conference on International Computing Education Research},
pages = {219–227},
numpages = {9},
location = {Toronto ON, Canada},
series = {ICER '19}
}

@article{McLaughlin_2025,
author = {McLaughlin, Gözde and Farris, Amy},
year = {2025},
month = {05},
pages = {1-25},
title = {A Systems Thinking Perspective on Building and Debugging Physical Computing Projects},
journal = {Technology, Knowledge and Learning},
doi = {10.1007/s10758-025-09855-5}
}

@article{Hennessy_2024,
author = {Hennessy Elliott, Colin and Nixon, Jessie and Gendrau Chakarov, Alexandra and Bush, Jeffrey B. and Schneider, Michael J. and Recker, Mimi},
title = {Characterizing Teacher Support of Debugging with Physical Computing: Debugging Pedagogies in Practice},
year = {2024},
issue_date = {December 2024},
publisher = {Association for Computing Machinery},
address = {New York, NY, USA},
volume = {24},
number = {4},
url = {https://doi.org/10.1145/3677612},
doi = {10.1145/3677612},
journal = {ACM Trans. Comput. Educ.},
month = dec,
articleno = {48},
numpages = {28}
}

@inproceedings{Torres_2019,
author = {Torres, Cesar and Nicholas, Molly Jane and Lee, Sangyeon and Paulos, Eric},
title = {A Conversation with Actuators: An Exploratory Design Environment for Hybrid Materials},
year = {2019},
isbn = {9781450361965},
publisher = {Association for Computing Machinery},
address = {New York, NY, USA},
url = {https://doi.org/10.1145/3294109.3295643},
doi = {10.1145/3294109.3295643},
booktitle = {Proceedings of the Thirteenth International Conference on Tangible, Embedded, and Embodied Interaction},
pages = {657–667},
numpages = {11},
location = {Tempe, Arizona, USA},
series = {TEI '19}
}

@InProceedings{TrustinAutomation,
author="K{\"o}rber, Moritz",
editor="Bagnara, Sebastiano
and Tartaglia, Riccardo
and Albolino, Sara
and Alexander, Thomas
and Fujita, Yushi",
title="Theoretical Considerations and Development of a Questionnaire to Measure Trust in Automation",
booktitle="Proceedings of the 20th Congress of the International Ergonomics Association (IEA 2018)",
year="2019",
publisher="Springer International Publishing",
address="Cham",
pages="13--30",
isbn="978-3-319-96074-6"
}

@inproceedings{Whyline,
author = {Ko, Amy J. and Myers, Brad A.},
title = {Designing the whyline: a debugging interface for asking questions about program behavior},
year = {2004},
isbn = {1581137028},
publisher = {Association for Computing Machinery},
address = {New York, NY, USA},
url = {https://doi.org/10.1145/985692.985712},
doi = {10.1145/985692.985712},
booktitle = {Proceedings of the SIGCHI Conference on Human Factors in Computing Systems},
pages = {151–158},
numpages = {8},
location = {Vienna, Austria},
series = {CHI '04}
}

@inproceedings{MakeDevice,
author = {Hartley, Kobi and Finney, Joe and Hodges, Steve and De Halleux, Peli and Devine, James and D'Amone, Gabriele},
title = {MakeDevice: Evolving Devices Beyond the Prototype with Jacdac},
year = {2023},
isbn = {9781450399777},
publisher = {Association for Computing Machinery},
address = {New York, NY, USA},
url = {https://doi.org/10.1145/3569009.3573106},
doi = {10.1145/3569009.3573106},
booktitle = {Proceedings of the Seventeenth International Conference on Tangible, Embedded, and Embodied Interaction},
articleno = {37},
numpages = {7},
location = {Warsaw, Poland},
series = {TEI '23}
}

@inproceedings{MeasurementPatterns,
author = {Ramakers, Raf and Leen, Danny and Kim, Jeeeun and Luyten, Kris and Houben, Steven and Veuskens, Tom},
title = {Measurement Patterns: User-Oriented Strategies for Dealing with Measurements and Dimensions in Making Processes},
year = {2023},
isbn = {9781450394215},
publisher = {Association for Computing Machinery},
address = {New York, NY, USA},
url = {https://doi.org/10.1145/3544548.3581157},
doi = {10.1145/3544548.3581157},
booktitle = {Proceedings of the 2023 CHI Conference on Human Factors in Computing Systems},
articleno = {214},
numpages = {17},
location = {Hamburg, Germany},
series = {CHI '23}
}

@inproceedings{Bianchi_2023,
author = {Bianchi, Andrea and Hodges, Steve and Cuartielles, David J. and Oh, Hyunjoo and Lambrichts, Mannu and Roudaut, Anne},
title = {Beyond prototyping boards: future paradigms for electronics toolkits},
year = {2023},
isbn = {9781450394222},
publisher = {Association for Computing Machinery},
address = {New York, NY, USA},
url = {https://doi.org/10.1145/3544549.3573792},
doi = {10.1145/3544549.3573792},
booktitle = {Extended Abstracts of the 2023 CHI Conference on Human Factors in Computing Systems},
articleno = {333},
numpages = {6},
location = {Hamburg, Germany},
series = {CHI EA '23}
}

@inproceedings{Lin_2021,
author = {Lin, Richard and Ramesh, Rohit and Jain, Nikhil and Koe, Josephine and Nuqui, Ryan and Dutta, Prabal and Hartmann, Bjoern},
title = {Weaving Schematics and Code: Interactive Visual Editing for Hardware Description Languages},
year = {2021},
isbn = {9781450386357},
publisher = {Association for Computing Machinery},
address = {New York, NY, USA},
url = {https://doi.org/10.1145/3472749.3474804},
doi = {10.1145/3472749.3474804},
booktitle = {The 34th Annual ACM Symposium on User Interface Software and Technology},
pages = {1039–1049},
numpages = {11},
location = {Virtual Event, USA},
series = {UIST '21}
}

@article{SUS_interpretation,
author = {Bangor, Aaron and Kortum, Philip and Miller, James},
title = {Determining what individual SUS scores mean: adding an adjective rating scale},
year = {2009},
issue_date = {May 2009},
publisher = {Usability Professionals' Association},
address = {Bloomingdale, IL},
volume = {4},
number = {3},
journal = {J. Usability Studies},
month = may,
pages = {114–123},
numpages = {10}
}

@inproceedings{CircuitStack,
author = {Wang, Chiuan and Yeh, Hsuan-Ming and Wang, Bryan and Wu, Te-Yen and Tsai, Hsin-Ruey and Liang, Rong-Hao and Hung, Yi-Ping and Chen, Mike Y.},
title = {CircuitStack: Supporting Rapid Prototyping and Evolution of Electronic Circuits},
year = {2016},
isbn = {9781450341899},
publisher = {Association for Computing Machinery},
address = {New York, NY, USA},
url = {https://doi.org/10.1145/2984511.2984527},
doi = {10.1145/2984511.2984527},
booktitle = {Proceedings of the 29th Annual Symposium on User Interface Software and Technology},
pages = {687–695},
numpages = {9},
location = {Tokyo, Japan},
series = {UIST '16}
}

@inproceedings{CurrentViz,
author = {Wu, Te-Yen and Shen, Hao-Ping and Wu, Yu-Chian and Chen, Yu-An and Ku, Pin-Sung and Hsu, Ming-Wei and Liu, Jun-You and Lin, Yu-Chih and Chen, Mike Y.},
title = {CurrentViz: Sensing and Visualizing Electric Current Flows of Breadboarded Circuits},
year = {2017},
isbn = {9781450349819},
publisher = {Association for Computing Machinery},
address = {New York, NY, USA},
url = {https://doi.org/10.1145/3126594.3126646},
doi = {10.1145/3126594.3126646},
booktitle = {Proceedings of the 30th Annual ACM Symposium on User Interface Software and Technology},
pages = {343–349},
numpages = {7},
location = {Qu\'{e}bec City, QC, Canada},
series = {UIST '17}
}

@article{BlinkBoard,
title = {BlinkBoard: Guiding and monitoring circuit assembly for synchronous and remote physical computing education},
journal = {HardwareX},
volume = {17},
pages = {e00511},
year = {2024},
issn = {2468-0672},
doi = {https://doi.org/10.1016/j.ohx.2024.e00511},
url = {https://www.sciencedirect.com/science/article/pii/S2468067224000051},
author = {Andrea Bianchi and Kongpyung (Justin) Moon and Artem Dementyev and Seungwoo Je}
}

@inproceedings{LogIt,
author = {Jiang, Peiling and Sun, Fuling and Xia, Haijun},
title = {Log-it: Supporting Programming with Interactive, Contextual, Structured, and Visual Logs},
year = {2023},
isbn = {9781450394215},
publisher = {Association for Computing Machinery},
address = {New York, NY, USA},
url = {https://doi.org/10.1145/3544548.3581403},
doi = {10.1145/3544548.3581403},
booktitle = {Proceedings of the 2023 CHI Conference on Human Factors in Computing Systems},
articleno = {594},
numpages = {16},
location = {Hamburg, Germany},
series = {CHI '23}
}

@inproceedings{Hartmann_2006,
author = {Hartmann, Bj\"{o}rn and Klemmer, Scott R. and Bernstein, Michael and Abdulla, Leith and Burr, Brandon and Robinson-Mosher, Avi and Gee, Jennifer},
title = {Reflective physical prototyping through integrated design, test, and analysis},
year = {2006},
isbn = {1595933131},
publisher = {Association for Computing Machinery},
address = {New York, NY, USA},
url = {https://doi.org/10.1145/1166253.1166300},
doi = {10.1145/1166253.1166300},
booktitle = {Proceedings of the 19th Annual ACM Symposium on User Interface Software and Technology},
pages = {299–308},
numpages = {10},
location = {Montreux, Switzerland},
series = {UIST '06}
}

@inproceedings{VisibleBreadboard,
author = {Ochiai, Yoichi},
title = {Visible Breadboard: System for Dynamic, Programmable, and Tangible Circuit Prototyping with Visible Electricity},
year = {2014},
isbn = {9783319074634},
publisher = {Springer-Verlag},
address = {Berlin, Heidelberg},
url = {https://doi.org/10.1007/978-3-319-07464-1_7},
doi = {10.1007/978-3-319-07464-1_7},
booktitle = {Proceedings of the 6th International Conference on Virtual, Augmented and Mixed Reality. Applications of Virtual and Augmented Reality - Volume 8526},
pages = {73–84},
numpages = {12}
}

@INPROCEEDINGS{Bajpai_2024,
  author={Bajpai, Yasharth and Chopra, Bhavya and Biyani, Param and Aslan, Cagri and Coleman, Dustin and Gulwani, Sumit and Parnin, Chris and Radhakrishna, Arjun and Soares, Gustavo},
  booktitle={2024 IEEE Symposium on Visual Languages and Human-Centric Computing (VL/HCC)}, 
  title={Let’s Fix this Together: Conversational Debugging with GitHub Copilot}, 
  year={2024},
  volume={},
  number={},
  pages={1-12},
  doi={10.1109/VL/HCC60511.2024.00011},
  ISSN={1943-6106},
  month={Sep.}}

@inproceedings{WiFrost,
author = {McGrath, William and Warner, Jeremy and Karchemsky, Mitchell and Head, Andrew and Drew, Daniel and Hartmann, Bjoern},
title = {WiFr\"{o}st: Bridging the Information Gap for Debugging of Networked Embedded Systems},
year = {2018},
isbn = {9781450359481},
publisher = {Association for Computing Machinery},
address = {New York, NY, USA},
url = {https://doi.org/10.1145/3242587.3242668},
doi = {10.1145/3242587.3242668},
booktitle = {Proceedings of the 31st Annual ACM Symposium on User Interface Software and Technology},
pages = {447–455},
numpages = {9},
location = {Berlin, Germany},
series = {UIST '18}
}

@inproceedings{PolymorphicBlocks,
author = {Lin, Richard and Ramesh, Rohit and Chi, Connie and Jain, Nikhil and Nuqui, Ryan and Dutta, Prabal and Hartmann, Bj\"{o}rn},
title = {Polymorphic Blocks: Unifying High-level Specification and Low-level Control for Circuit Board Design},
year = {2020},
isbn = {9781450375146},
publisher = {Association for Computing Machinery},
address = {New York, NY, USA},
url = {https://doi.org/10.1145/3379337.3415860},
doi = {10.1145/3379337.3415860},
booktitle = {Proceedings of the 33rd Annual ACM Symposium on User Interface Software and Technology},
pages = {529–540},
numpages = {12},
location = {Virtual Event, USA},
series = {UIST '20}
}

@article{FritzBot,
title = {FritzBot: A data-driven conversational agent for physical-computing system design},
journal = {International Journal of Human-Computer Studies},
volume = {155},
pages = {102699},
year = {2021},
issn = {1071-5819},
doi = {https://doi.org/10.1016/j.ijhcs.2021.102699},
url = {https://www.sciencedirect.com/science/article/pii/S1071581921001178},
author = {Taizhou Chen and Lantian Xu and Kening Zhu}
}

@inproceedings{Lin_2019,
author = {Lin, Richard and Ramesh, Rohit and Iannopollo, Antonio and Sangiovanni Vincentelli, Alberto and Dutta, Prabal and Alon, Elad and Hartmann, Bj\"{o}rn},
title = {Beyond Schematic Capture: Meaningful Abstractions for Better Electronics Design Tools},
year = {2019},
isbn = {9781450359702},
publisher = {Association for Computing Machinery},
address = {New York, NY, USA},
url = {https://doi.org/10.1145/3290605.3300513},
doi = {10.1145/3290605.3300513},
booktitle = {Proceedings of the 2019 CHI Conference on Human Factors in Computing Systems},
pages = {1–13},
numpages = {13},
location = {Glasgow, Scotland Uk},
series = {CHI '19}
}

@inproceedings{SensorViz,
author = {Kim, Yoonji and Zhu, Junyi and Trivedi, Mihir and Turakhia, Dishita and Wu, Ngai Hang and Ko, Donghyeon and Wessely, Michael and Mueller, Stefanie},
title = {SensorViz: Visualizing Sensor Data Across Different Stages of Prototyping Interactive Objects},
year = {2022},
isbn = {9781450393584},
publisher = {Association for Computing Machinery},
address = {New York, NY, USA},
url = {https://doi.org/10.1145/3532106.3533481},
doi = {10.1145/3532106.3533481},
booktitle = {Proceedings of the 2022 ACM Designing Interactive Systems Conference},
pages = {987–1001},
numpages = {15},
location = {Virtual Event, Australia},
series = {DIS '22}
}

@inproceedings{VirtualWire,
author = {Lee, Woojin and Prasad, Ramkrishna and Je, Seungwoo and Kim, Yoonji and Oakley, Ian and Ashbrook, Daniel and Bianchi, Andrea},
title = {VirtualWire: Supporting Rapid Prototyping with Instant Reconfigurations of Wires in Breadboarded Circuits},
year = {2021},
isbn = {9781450382137},
publisher = {Association for Computing Machinery},
address = {New York, NY, USA},
url = {https://doi.org/10.1145/3430524.3440623},
doi = {10.1145/3430524.3440623},
booktitle = {Proceedings of the Fifteenth International Conference on Tangible, Embedded, and Embodied Interaction},
articleno = {4},
numpages = {12},
location = {Salzburg, Austria},
series = {TEI '21}
}

@inproceedings{Proxino,
author = {Wu, Te-Yen and Gong, Jun and Seyed, Teddy and Yang, Xing-Dong},
title = {Proxino: Enabling Prototyping of Virtual Circuits with Physical Proxies},
year = {2019},
isbn = {9781450368162},
publisher = {Association for Computing Machinery},
address = {New York, NY, USA},
url = {https://doi.org/10.1145/3332165.3347938},
doi = {10.1145/3332165.3347938},
booktitle = {Proceedings of the 32nd Annual ACM Symposium on User Interface Software and Technology},
pages = {121–132},
numpages = {12},
location = {New Orleans, LA, USA},
series = {UIST '19}
}

@inproceedings{Yang_2023,
author = {Yang, Saelyne and Kwak, Sangkyung and Lee, Juhoon and Kim, Juho},
title = {Beyond Instructions: A Taxonomy of Information Types in How-to Videos},
year = {2023},
isbn = {9781450394215},
publisher = {Association for Computing Machinery},
address = {New York, NY, USA},
url = {https://doi.org/10.1145/3544548.3581126},
doi = {10.1145/3544548.3581126},
booktitle = {Proceedings of the 2023 CHI Conference on Human Factors in Computing Systems},
articleno = {797},
numpages = {21},
location = {Hamburg, Germany},
series = {CHI '23}
}

@inproceedings{Heimdall,
author = {Karchemsky, Mitchell and Zamfirescu-Pereira, J.D. and Wu, Kuan-Ju and Guimbreti\`{e}re, Fran\c{c}ois and Hartmann, Bjoern},
title = {Heimdall: A Remotely Controlled Inspection Workbench For Debugging Microcontroller Projects},
year = {2019},
isbn = {9781450359702},
publisher = {Association for Computing Machinery},
address = {New York, NY, USA},
url = {https://doi.org/10.1145/3290605.3300728},
doi = {10.1145/3290605.3300728},
booktitle = {Proceedings of the 2019 CHI Conference on Human Factors in Computing Systems},
pages = {1–12},
numpages = {12},
location = {Glasgow, Scotland Uk},
series = {CHI '19}
}

@article{Inline,
author = {Bianchi, Andrea and Yap, Zhi Lin and Lertjaturaphat, Punn and Henley, Austin Z. and Moon, Kongpyung Justin and Kim, Yoonji},
title = {Inline Visualization and Manipulation of Real-Time Hardware Log for Supporting Debugging of Embedded Programs},
year = {2024},
issue_date = {June 2024},
publisher = {Association for Computing Machinery},
address = {New York, NY, USA},
volume = {8},
number = {EICS},
url = {https://doi.org/10.1145/3660250},
doi = {10.1145/3660250},
journal = {Proc. ACM Hum.-Comput. Interact.},
month = jun,
articleno = {248},
numpages = {26}
}

@inproceedings{Ko_2008,
author = {Ko, Amy J. and Myers, Brad A.},
title = {Debugging reinvented: asking and answering why and why not questions about program behavior},
year = {2008},
isbn = {9781605580791},
publisher = {Association for Computing Machinery},
address = {New York, NY, USA},
url = {https://doi.org/10.1145/1368088.1368130},
doi = {10.1145/1368088.1368130},
booktitle = {Proceedings of the 30th International Conference on Software Engineering},
pages = {301–310},
numpages = {10},
location = {Leipzig, Germany},
series = {ICSE '08}
}

@inproceedings{VirtualComponent,
author = {Kim, Yoonji and Choi, Youngkyung and Lee, Hyein and Lee, Geehyuk and Bianchi, Andrea},
title = {VirtualComponent: A Mixed-Reality Tool for Designing and Tuning Breadboarded Circuits},
year = {2019},
isbn = {9781450359702},
publisher = {Association for Computing Machinery},
address = {New York, NY, USA},
url = {https://doi.org/10.1145/3290605.3300407},
doi = {10.1145/3290605.3300407},
booktitle = {Proceedings of the 2019 CHI Conference on Human Factors in Computing Systems},
pages = {1–13},
numpages = {13},
location = {Glasgow, Scotland Uk},
series = {CHI '19}
}

@book{fitzgerald2012arduino,
  title={Arduino Projects Book},
  author={Fitzgerald, S. and Shiloh, M. and Igoe, T.},
  url={https://books.google.co.kr/books?id=4Bxz0AEACAAJ},
  year={2012},
  publisher={Arduino}
}

@inproceedings{Toastboard,
author = {Drew, Daniel and Newcomb, Julie L. and McGrath, William and Maksimovic, Filip and Mellis, David and Hartmann, Bj\"{o}rn},
title = {The Toastboard: Ubiquitous Instrumentation and Automated Checking of Breadboarded Circuits},
year = {2016},
isbn = {9781450341899},
publisher = {Association for Computing Machinery},
address = {New York, NY, USA},
url = {https://doi.org/10.1145/2984511.2984566},
doi = {10.1145/2984511.2984566},
booktitle = {Proceedings of the 29th Annual Symposium on User Interface Software and Technology},
pages = {677–686},
numpages = {10},
location = {Tokyo, Japan},
series = {UIST '16}
}

@inproceedings{Scanalog,
author = {Strasnick, Evan and Agrawala, Maneesh and Follmer, Sean},
title = {Scanalog: Interactive Design and Debugging of Analog Circuits with Programmable Hardware},
year = {2017},
isbn = {9781450349819},
publisher = {Association for Computing Machinery},
address = {New York, NY, USA},
url = {https://doi.org/10.1145/3126594.3126618},
doi = {10.1145/3126594.3126618},
booktitle = {Proceedings of the 30th Annual ACM Symposium on User Interface Software and Technology},
pages = {321–330},
numpages = {10},
location = {Qu\'{e}bec City, QC, Canada},
series = {UIST '17}
}

@inproceedings{CircuitStyle,
author = {Davis, Josh Urban and Gong, Jun and Sun, Yunxin and Chilana, Parmit and Yang, Xing-Dong},
title = {CircuitStyle: A System for Peripherally Reinforcing Best Practices in Hardware Computing},
year = {2019},
isbn = {9781450368162},
publisher = {Association for Computing Machinery},
address = {New York, NY, USA},
url = {https://doi.org/10.1145/3332165.3347920},
doi = {10.1145/3332165.3347920},
booktitle = {Proceedings of the 32nd Annual ACM Symposium on User Interface Software and Technology},
pages = {109–120},
numpages = {12},
location = {New Orleans, LA, USA},
series = {UIST '19}
}

@inproceedings{SchemaBoard,
author = {Kim, Yoonji and Lee, Hyein and Prasad, Ramkrishna and Je, Seungwoo and Choi, Youngkyung and Ashbrook, Daniel and Oakley, Ian and Bianchi, Andrea},
title = {SchemaBoard: Supporting Correct Assembly of Schematic Circuits using Dynamic In-Situ Visualization},
year = {2020},
isbn = {9781450375146},
publisher = {Association for Computing Machinery},
address = {New York, NY, USA},
url = {https://doi.org/10.1145/3379337.3415887},
doi = {10.1145/3379337.3415887},
booktitle = {Proceedings of the 33rd Annual ACM Symposium on User Interface Software and Technology},
pages = {987–998},
numpages = {12},
location = {Virtual Event, USA},
series = {UIST '20}
}

@inproceedings{Bifrost,
author = {McGrath, Will and Drew, Daniel and Warner, Jeremy and Kazemitabaar, Majeed and Karchemsky, Mitchell and Mellis, David and Hartmann, Bj\"{o}rn},
title = {Bifr\"{o}st: Visualizing and Checking Behavior of Embedded Systems across Hardware and Software},
year = {2017},
isbn = {9781450349819},
publisher = {Association for Computing Machinery},
address = {New York, NY, USA},
url = {https://doi.org/10.1145/3126594.3126658},
doi = {10.1145/3126594.3126658},
booktitle = {Proceedings of the 30th Annual ACM Symposium on User Interface Software and Technology},
pages = {299–310},
numpages = {12},
location = {Qu\'{e}bec City, QC, Canada},
series = {UIST '17}
}

@inproceedings{CircuitSense,
author = {Wu, Te-Yen and Wang, Bryan and Lee, Jiun-Yu and Shen, Hao-Ping and Wu, Yu-Chian and Chen, Yu-An and Ku, Pin-Sung and Hsu, Ming-Wei and Lin, Yu-Chih and Chen, Mike Y.},
title = {CircuitSense: Automatic Sensing of Physical Circuits and Generation of Virtual Circuits to Support Software Tools.},
year = {2017},
isbn = {9781450349819},
publisher = {Association for Computing Machinery},
address = {New York, NY, USA},
url = {https://doi.org/10.1145/3126594.3126634},
doi = {10.1145/3126594.3126634},
booktitle = {Proceedings of the 30th Annual ACM Symposium on User Interface Software and Technology},
pages = {311–319},
numpages = {9},
location = {Qu\'{e}bec City, QC, Canada},
series = {UIST '17}
}

@inproceedings{MixT,
author = {Chi, Pei-Yu and Ahn, Sally and Ren, Amanda and Dontcheva, Mira and Li, Wilmot and Hartmann, Bj\"{o}rn},
title = {MixT: automatic generation of step-by-step mixed media tutorials},
year = {2012},
isbn = {9781450315807},
publisher = {Association for Computing Machinery},
address = {New York, NY, USA},
url = {https://doi.org/10.1145/2380116.2380130},
doi = {10.1145/2380116.2380130},
booktitle = {Proceedings of the 25th Annual ACM Symposium on User Interface Software and Technology},
pages = {93–102},
numpages = {10},
location = {Cambridge, Massachusetts, USA},
series = {UIST '12}
}

@article{HeyTeddy,
author = {Kim, Yoonji and Choi, Youngkyung and Kang, Daye and Lee, Minkyeong and Nam, Tek-Jin and Bianchi, Andrea},
title = {HeyTeddy: Conversational Test-Driven Development for Physical Computing},
year = {2020},
issue_date = {December 2019},
publisher = {Association for Computing Machinery},
address = {New York, NY, USA},
volume = {3},
number = {4},
url = {https://doi.org/10.1145/3369838},
doi = {10.1145/3369838},
journal = {Proc. ACM Interact. Mob. Wearable Ubiquitous Technol.},
month = sep,
articleno = {139},
numpages = {21}
}

@inproceedings{ElectroTutor,
author = {Warner, Jeremy and Lafreniere, Ben and Fitzmaurice, George and Grossman, Tovi},
title = {ElectroTutor: Test-Driven Physical Computing Tutorials},
year = {2018},
isbn = {9781450359481},
publisher = {Association for Computing Machinery},
address = {New York, NY, USA},
url = {https://doi.org/10.1145/3242587.3242591},
doi = {10.1145/3242587.3242591},
booktitle = {Proceedings of the 31st Annual ACM Symposium on User Interface Software and Technology},
pages = {435–446},
numpages = {12},
location = {Berlin, Germany},
series = {UIST '18}
}

@article{Flowboard,
author = {Brocker, Anke and Sch\"{a}fer, Ren\'{e} and Remy, Christian and Voelker, Simon and Borchers, Jan},
title = {Flowboard: How Seamless, Live, Flow-Based Programming Impacts Learning to Code for Embedded Electronics},
year = {2023},
issue_date = {February 2023},
publisher = {Association for Computing Machinery},
address = {New York, NY, USA},
volume = {30},
number = {1},
issn = {1073-0516},
url = {https://doi.org/10.1145/3533015},
doi = {10.1145/3533015},
journal = {ACM Trans. Comput.-Hum. Interact.},
month = mar,
articleno = {2},
numpages = {36}
}

@inproceedings{Fritzing,
author = {Kn\"{o}rig, Andr\'{e} and Wettach, Reto and Cohen, Jonathan},
title = {Fritzing: a tool for advancing electronic prototyping for designers},
year = {2009},
isbn = {9781605584935},
publisher = {Association for Computing Machinery},
address = {New York, NY, USA},
url = {https://doi.org/10.1145/1517664.1517735},
doi = {10.1145/1517664.1517735},
booktitle = {Proceedings of the 3rd International Conference on Tangible and Embedded Interaction},
pages = {351–358},
numpages = {8},
location = {Cambridge, United Kingdom},
series = {TEI '09}
}

@inproceedings{AutoFritz,
author = {Lo, Jo-Yu and Huang, Da-Yuan and Kuo, Tzu-Sheng and Sun, Chen-Kuo and Gong, Jun and Seyed, Teddy and Yang, Xing-Dong and Chen, Bing-Yu},
title = {AutoFritz: Autocomplete for Prototyping Virtual Breadboard Circuits},
year = {2019},
isbn = {9781450359702},
publisher = {Association for Computing Machinery},
address = {New York, NY, USA},
url = {https://doi.org/10.1145/3290605.3300633},
doi = {10.1145/3290605.3300633},
booktitle = {Proceedings of the 2019 CHI Conference on Human Factors in Computing Systems},
pages = {1–13},
numpages = {13},
location = {Glasgow, Scotland Uk},
series = {CHI '19}
}

@inproceedings{TriggerActionCircuits,
author = {Anderson, Fraser and Grossman, Tovi and Fitzmaurice, George},
title = {Trigger-Action-Circuits: Leveraging Generative Design to Enable Novices to Design and Build Circuitry},
year = {2017},
isbn = {9781450349819},
publisher = {Association for Computing Machinery},
address = {New York, NY, USA},
url = {https://doi.org/10.1145/3126594.3126637},
doi = {10.1145/3126594.3126637},
booktitle = {Proceedings of the 30th Annual ACM Symposium on User Interface Software and Technology},
pages = {331–342},
numpages = {12},
location = {Qu\'{e}bec City, QC, Canada},
series = {UIST '17}
}

@inproceedings{CrossedWires,
author = {Booth, Tracey and Stumpf, Simone and Bird, Jon and Jones, Sara},
title = {Crossed Wires: Investigating the Problems of End-User Developers in a Physical Computing Task},
year = {2016},
isbn = {9781450333627},
publisher = {Association for Computing Machinery},
address = {New York, NY, USA},
url = {https://doi.org/10.1145/2858036.2858533},
doi = {10.1145/2858036.2858533},
booktitle = {Proceedings of the 2016 CHI Conference on Human Factors in Computing Systems},
pages = {3485–3497},
numpages = {13},
location = {San Jose, California, USA},
series = {CHI '16}
}

@article{GeneralInductive,
author = {David R. Thomas},
title ={A General Inductive Approach for Analyzing Qualitative Evaluation Data},

journal = {American Journal of Evaluation},
volume = {27},
number = {2},
pages = {237-246},
year = {2006},
doi = {10.1177/1098214005283748},

URL = { 
    
        https://doi.org/10.1177/1098214005283748
    
    

},
eprint = { 
    
        https://doi.org/10.1177/1098214005283748
    
    

}
}

\appendix

\section{Formative Study: Interview Protocol} \label{appendix:formative}

We include here the full set of interview questions used in the formative study.
\begin{enumerate}
    \item How did you decide on the keywords or sites you used?
    \item Were there any search results you ignored right away? Why?
    \item Which media format of information did you go to first, and why?
    \item Were there moments you switched formats? What triggered the switch?
    \item How did you judge the trustworthiness or accuracy of the online tutorials? Can you give an example?
    \item Did you find any format misleading or confusing? What made it so?
    \item How did you keep track of different wiring diagrams or code fragments you encountered?
    \item Did you feel like you were jumping between multiple tutorials? Tell me about that experience.
    \item When you found a snippet that looked useful, how did you test or verify it before applying it?
    \item Tell me about one mistake you made (wiring vs.\ code) and how you diagnosed or fixed it.
    \item How would you redesign the tutorial to avoid the mistake you explained?
    \item Would you wish there were any hints or checkpoints that could have helped?
    \item At any point, did you wish all the information were in one place? What would that look like?
    \item Anything else you wish we had asked or that you’d like to share?
\end{enumerate}

\section{System Interface} \label{appendix:system_interface}
\begin{figure}[h!]
    \centering
    \includegraphics[width=1\linewidth]{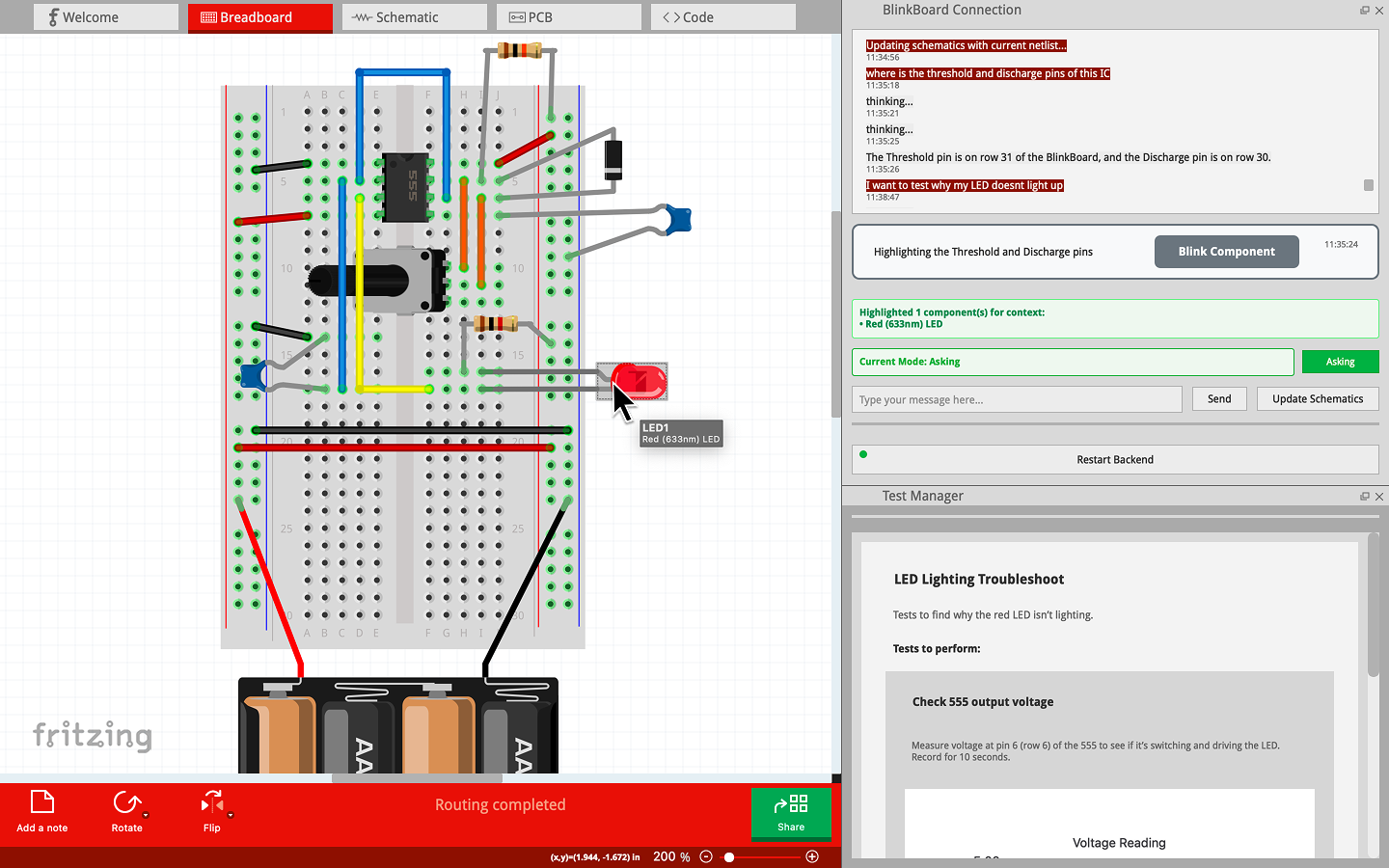}
    \caption{Unmodified user interface of \systemName}
    \label{fig:unmodified_interface}
    \Description{Three-panel interface layout. Left panel: White breadboard populated with electronic components including 555 timer IC, potentiometer, resistors, diodes, red LED (LED1) with tooltip under cursor, blue LED, and colored wires connecting components. Brown three-AA battery pack connected at bottom. Bottom bar shows "Routing completed" and "200\%" zoom. Top-right panel: "BlinkBoard Connection" chat interface displaying conversation history with red user queries and system responses, "Blink Component" button, green banner showing highlighted component context, and "Asking" mode button with text input field. Bottom-right panel: "Test Manager" showing "LED Lighting Troubleshoot" with test list, detailed instructions for "Check 555 output voltage" test, and empty "Voltage Reading" area.}
\end{figure}

\section{User Study: Evaluation Questionnaires} \label{appendix:evaluationq}
Participants rated the following statements on a 5-point Likert scale. 
(\textit{1 = Strongly Disagree, 5 = Strongly Agree}).

\begin{enumerate}
    \item The tests helped me isolate faults quickly. \hfill (1--5)
    \item Measurement suggestions matched what mattered on the board. \hfill (1--5)
    \item Test outcomes mapped clearly to what I should do next. \hfill (1--5)
    \item I could tell when the test was wrong or inconclusive. \hfill (1--5)
    \item I understood why the system suggested each step. \hfill (1--5)
    \item When results changed, I knew what caused the change. \hfill (1--5)
    \item The link between schematic and board actions was clear. \hfill (1--5)
\end{enumerate}

\section{User Study: Interview Protocol} \label{appendix:userstudy}

We include here the full set of interview questions used in the usability study.

\subsection{Experience with AI for Physical Computing}
\begin{enumerate}
    \item Do you have experience in using any AI for physical computing? What was it like?
    \item What would be different if you were using that AI compared to our system?
\end{enumerate}

\subsection{System Usage and Information Seeking}
\begin{enumerate}
    \setcounter{enumi}{2}
    \item Were there any moments you wanted to search online for anything else?
    \item What was the most helpful aspect of the system?
    \item What was the most frustrating aspect of the system?
\end{enumerate}

\subsection{System Improvement and Future Use}
\begin{enumerate}
    \setcounter{enumi}{5}
    \item If you could change one thing about the system, what would it be?
    \item Can you imagine using this system in any context? What projects would you use it for, and what projects would you not use it for?
\end{enumerate}

\subsection{Task-Specific Clarifications}
Additional clarification questions were asked based on observed participant behaviors and task-specific challenges encountered during the session.
\end{document}